\documentclass[a4paper, 11pt]{article}
\pdfoutput=1
\usepackage{jheppub}
\usepackage{subcaption}
\usepackage{amsmath}
\usepackage{amssymb}
\usepackage{amscd}
\usepackage{dsfont}
\usepackage{enumerate}
\usepackage{enumitem}
\usepackage{amsfonts}
\usepackage{epsfig}
\usepackage{mathtools}
\usepackage{yfonts}
\usepackage{bbold}
\usepackage{cancel}
\usepackage{breqn}
\usepackage{natbib}
\usepackage[utf8]{inputenc}
\usepackage[english]{babel}

\def\bal#1\eal{\begin{align}#1\end{align}}
\def\alp[#1]{\begin{align}#1\end{align}}
\def\secnum[#1]{\texorpdfstring{$#1$}{TEXT}}
\def\secnuml#1\secnumr{\texorpdfstring{$#1$}{TEXT}}

\def\eqa{\begin{eqnarray}}
    \def\eqae{\end{eqnarray}}
\def\eq{\begin{equation}}
    \def\eqe{\end{equation}}
\def\be{\begin{equation}}
    \def\ee{\end{equation}}
\def\bea{\begin{eqnarray}}
    \def\eea{\end{eqnarray}}
\def\ba{\begin{array}}
    \def\ea{\end{array}}
\def\bd{\begin{displaymath}}
    \def\ed{\end{displaymath}}

\def\ie{{\it i.e.~}}

\def\>{\rangle}
\def\<{\langle}

\title{Shape dependence of mutual information in the OPE limit: linear responses}
\begin{document}
\author{Liangyu Chen}
\author{and Huajia Wang}
\affiliation{Kavli Institute for Theoretical Sciences, University of Chinese Academy of Sciences, \\ Beijing 100190, China}
\emailAdd{chenliangyu19@mails.ucas.ac.cn}
\emailAdd{wanghuajia@ucas.ac.cn}

\abstract{Mutual information serves as an important measure of correlation between subsystem components. In the framework of quantum field theories (QFTs) they have better regulated UV behavior than entanglement entropy, and thus provide more direct access to universal aspects of entanglement structures. In this paper, we study the linear responses under shape deformation of the mutual information in the conformal field theory (CFT) vacuum between two spheres of radius $R$ separated by large distance $L\gg R$ or conformally equivalent configurations. Our calculations make use of the previous OPE results for mutual information \cite{Faulkner2016Aug} and the associated modular Hamiltonian \cite{Faulkner2021Aug}. In particular, we apply the entanglement  first law to compute the linear responses of mutual information under shape deformation on one of the spheres. We find that the linear responses exhibit a high degree of universality for a selected class of OPE contributions. We demonstrate that there is a ``little group" of symmetries associated with the set-up. Our result implies that the spherical mutual information is extremal over shape deformations of non-zero modes under the symmetry group. }

\maketitle

\setcounter{page}{1}
\setcounter{tocdepth}{1}

\tableofcontents
\section{Introduction}\label{sec: intro}
Entanglement structure has played a key role in probing deep aspects of quantum field theories (QFTs) not accessible using conventional tools and techniques. For example, it can provide the order parameter for topological phase \cite{Kitaev2005Mar,Xiao-GangMar}; through the famous Ryu-Takayanagi formula \cite{Ryu2006} and its subsequent generalizations \cite{Hubeny2007}, it encodes information regarding bulk space-time geometry in holography \cite{Maldacena1999,Witten1998AntideSS, GUBSER1998105}; it can be used to characterize universal monotonous properties of renormalization group flow \cite{Myers2011Jan, Casini2011May,Casini:2015woa}; and it is also crucial for revealing connection between information theory and energy conditions \cite{HWang2016Sep, HWang2019Sep, Faulkner1804}, etc.

In principle, one can define various measures of the underlying entanglement structure. An important one is the entanglement entropy between subsystem $A$ and its complement $\bar{A}$. For a global pure state $|\psi\rangle$, the entanglement entropy is defined by:
\be
S^{\psi}_A \equiv -\text{tr}_A \rho^\psi_A \ln{\rho^\psi_A},\;\;\rho^\psi_A = \text{tr}_{\bar{A}}|\psi\rangle \langle \psi|
\ee
This definition relies on the notion of reduced density matrix $\rho^\psi_A$, which only makes clear sense if the global Hilbert space factorizes into tensor product $\mathcal{H}=\mathcal{H}_A \otimes \mathcal{H}_{\bar{A}}$. This is a convenient but usually problematic assumption in quantum field theories, for example due to gauge constraints imposed at the entangling boundary $\partial A$ \cite{William2012, Casini2014Apr, Lin:2018}. Another aspect associated with $\partial A$ in quantum field theories is the ubiquitous ultra-violet (UV) divergences that arise when computing entanglement entropies. Roughly speaking, they come from short-range entanglements across $\partial A$ that are only regularized by the UV cut-off of the theory. For example, in CFT the divergences take the form:
\be
S=c_{d-2}\frac{R^{d-2}}{\delta^{d-2}} + c_{d-4}\frac{R^{d-4}}{\delta^{d-4}} + ...
\begin{Bmatrix}
&+& c_{-1},\;\;\;d=\text{ odd}\\
&+& c_0 \log{\left(\frac{R}{\delta}\right)},\;d=\text{ even}
\end{Bmatrix}
\ee
where $R, \delta$ is the scale of entangling region and UV cutoff respectively, $c_{d-2}, c_{d-4},...$ are coefficients that depend on the geometry of entangling regions, and $c_{-1}$ and $c_0$ are universal. As a result, additional efforts are required in order to extract universal aspects of entanglement structures from the result of entanglement entropy. In order to do this, one usually construct combinations of entanglement quantities whose UV dependences cancel out. In practice, such cancellations can be arranged in a number of ways. For example, when extracting the order parameters for topological phases in 2+1 dimensional gapped systems, one computes the so-called topological entanglement entropy \cite{Kitaev2005Mar,Xiao-GangMar} by adding and subtracting vacuum entanglement entropies associated with different regions:
\be
 S_{\text{topo}} \equiv S_A +S_B+S_C-S_{AB}-S_{BC}-S_{AC}+S_{ABC}
\ee
out of which the shape dependence of these regions cancel out completely and leaving only topological contributions. Fixing a particular entangling region $A$, one can also arrange the cancellation between states, for example when computing the so-called relative entropy between two global states $\psi$ and $\sigma$:
\be
S_A\left(\psi|\sigma\right) \equiv \text{tr} \rho^\psi_A \ln{\rho^\psi_A} - \text{tr}\rho^\psi_A \ln{\rho^\sigma_A}
\ee
The relative entropy $S_A\left(\psi|\sigma\right)$ provides a measure of distinguishability between the global states $\psi$ and $\sigma$ based on what one can access in the subsystem $A$, i.e. between $\rho^\psi_A$ and $\rho^\sigma_A$. It satisfies several constraints such as positivity and monotonicity, and when applied to specific context they can sometimes reveal important physics such as the emergence of Einstein's equation in AdS/CFT \cite{Faulkner2014Mar, Faulkner2015May, Faulkner2017Aug, Lewkowycz2018} or the validity of the averaged null energy condition (ANEC) in QFTs \cite{HWang2016Sep}, etc.

Mutual information is one of the simplest combinations among these entanglement quantities. For disjoint entangling regions $A$ and $B$, one can define the mutual information $I_{A,B}$ between them as:
\be
I_{A,B} \equiv S_{A\cup B}-S_A-S_B
\ee
where we have omitted the dependence on the state $\psi$ in these quantities, since in this paper shall focus on the vacuum state $|\psi\rangle = |\Omega\rangle$ from now on. The UV divergences associated with short-distance entanglements near $\partial \left(A\cup B\right)$ are cancelled in $I_{A,B}$, thus making it a universal quantity. Roughly speaking $I_{AB}$ measures the correlation between the subsystems $A$ and $B$. When we view $A\cup B$ together as a system in the mixed state $\rho_{AB}$, $I_{A,B}$ receives contributions from both classical and quantum correlations. In attempts to refine the distinction between classical and quantum correlations in mixed states, other measures have been proposed which include entanglement negativity \cite{Eisert1999, Vidal2002, Plenio2005}. Being a universal quantity, the mutual information behaves in ways that align more with intuitions. For example due to strong subadditivity (SSA) for entanglement entropies, the mutual information like relative entropies also satisfy monotonicity, i.e.:
\be
I_{A,B}\leq I_{\tilde{A},B}
\ee
for $A \subseteq \tilde{A}$, while the entanglement entropies in general do not satisfy such constraints by themselves. On the other hand, the mutual information is less constrained by symmetries and thus encode more detailed information about the underlying theories. For example, while the single-interval vacuum entanglement entropies in 2d CFT ares constrained by symmetries to only depend on the central charges $c$, the mutual information between single-intervals depend on the full operator-spectrum of the CFTs.

Apart from the choice of entanglement measures, shape dependence on the entangling region can also reveal important information regarding entanglement structure. For example, in the presence of corner or cusp the entanglement entropy receives log divergent contributions whose coefficients on the other hand contain universal information \cite{Fradkin2006,Casini2007Dec,Hirata2007}. For regions with smooth boundaries, more generally one can study shape dependence by computing perturbation theory under shape deformation \cite{Rosenhaus2014Dec, Rosenhaus2015Feb}, usually about symmetric shapes i.e. spheres, half-planes etc. For spheres in CFTs, computing the first-order response indicates that its entanglement entropy is extremal under symmetry-breaking deformations \cite{Mezei2015Feb, Mezei2015Feb02}; while it being perturbatively minimal was verified by the second-order response which also revealed universal non-local contributions known as the entanglement density \cite{Faulkner2016Apr}; the minimization was later proven beyond perturbation theory \cite{Bueno2021}. Sometimes by probing shape dependence of carefully chosen entanglement quantities, one discover surprising connections between entanglement behaviors and other aspects of QFTs, such as the irreversibility of RG flows  and the validity of energy conditions, manifested through analyzing shape dependence of entanglement entropies (appropriately regularized) and relative entropies.

\begin{figure}[h]
    \centering
    \includegraphics[scale=0.52]{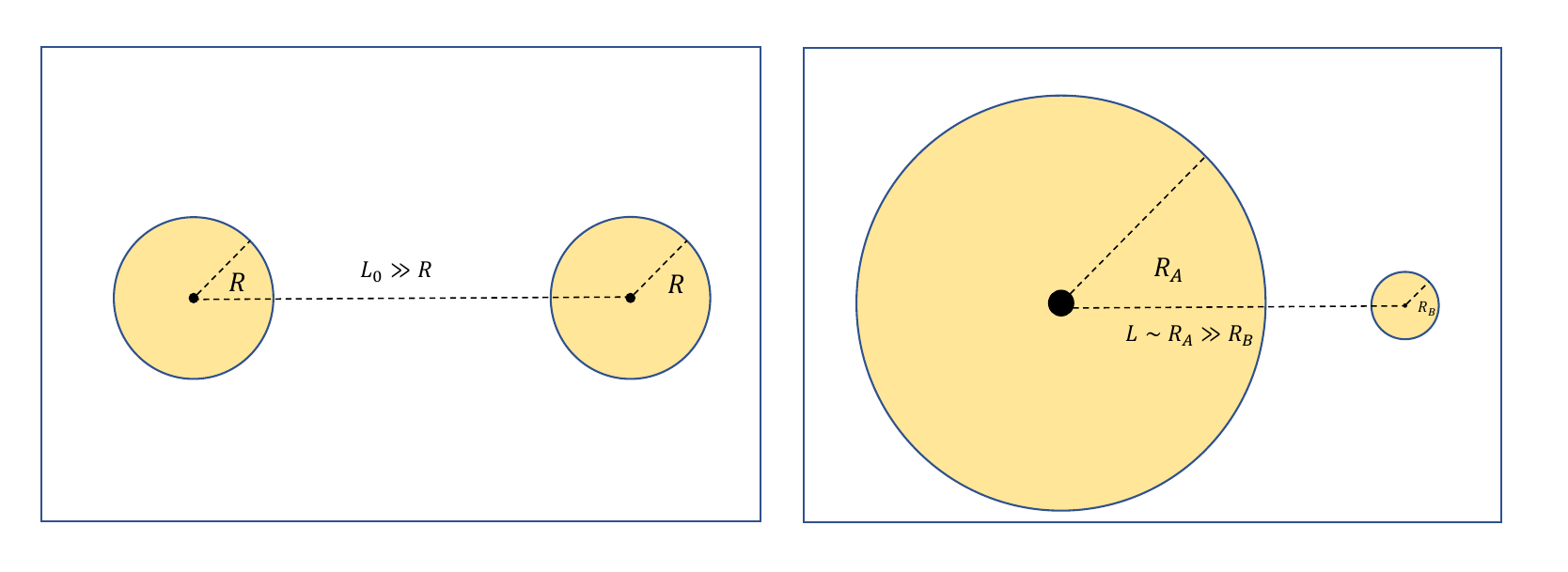}
    \caption{Two conformally equivalent configurations for the mutual information. The left: two spherical subregions of radius $R $ separated by a distance $L_{0} \gg R $. The right: two spherical subregions of radius $R_{A}$ and $R_B$ separated with distance $L  \sim R_{A} \gg R_{B}$. }
    \label{two-equivalent-frame-setup}
\end{figure}

It is therefore natural to expect that the shape dependence of mutual information can also reflect important physics. For example, it has been shown that the shape dependence of $I_{A,B}$ satisfies constraints on the null cone of $A$ which in the large separation limit (see \ref{eq:MI_OPE} below) between $A$ and $B$ can give rise to the unitarity bound \cite{Casini2021Sep}.  However, explicit results for shape dependence of mutual information using the general perturbative approach have been missing so far. One reason for this is that we do not have configurations with non-trivial mutual information that are sufficiently symmetric. In the case of a single sphere $A$ the vacuum entanglement structure in CFT is fixed by symmetry. In particular the modular Hamiltonian which encodes the complete entanglement data is known explicitly \cite{Casini2011May}:
\be
\hat{H}_A = 2 \pi \int_A dx \left(\frac{R^2-|x^2|}{2R}\right)\hat{T}_{tt}(x),\;\;\rho_A = e^{-\hat{H}_A}
\ee
Therefore it makes sense to do perturbation theory about it as in \cite{Mezei2015Feb,Mezei2015Feb02,Faulkner2016Apr}. On the other hand, the mutual information $I_{A,B}$ between two spheres $A$ and $B$ in CFT vacuum is not known in general form, despite being the most symmetric configuration with non-trivial mutual information. As mentioned before it is sensitive to the details (e.g. operator spectrum) of the theory. Explicit results can be obtained by restricting to special theories, e.g. Dirac fermions in two-dimensions \cite{Casini2005Jul,Casini2009Mar,Casini2009Sep}, in which the shape dependence on entangling region is degenerate; or by restricting to special limits in general CFTs, e.g. of large separation $L$ between $A$ and $B$ (see Fig \ref{two-equivalent-frame-setup}), where $I_{AB}$ admits an OPE-like expansion \cite{Cardy2013, Faulkner2016Aug, ChenBin2017, Long2016}:
\be\label{eq:MI_OPE}
I_{A,B}  =\mathcal{N}_{\Delta} \frac{\sqrt{\pi} \Gamma(2 \Delta+1)}{4 \Gamma\left(2 \Delta+\frac{3}{2}\right)} \left(\frac{R_{A} R_{B}}{L^2}\right)^{2 \Delta}+\ldots
\ee
where $\Delta$ denotes the conformal dimension of an internal scalar primary operator that ``carries" the correlation between $A$ and $B$ in some loose sense. The OPE result (\ref{eq:MI_OPE}) is valid in any dimensional CFTs, and therefore provides a good starting point to develop perturbative expansion in shape deformations.

In this paper, we study the shape dependence of mutual information perturbatively about spheres. We will focus on the leading order contribution in the OPE (\ref{eq:MI_OPE}). More specifically, we will compute the first-order linear response under shape deformation about one of the spheres $A$. The OPE expansion (\ref{eq:MI_OPE}) is conformally invariant, i.e. depending only on the conformal-ratio:
\be
\rho =\frac{2R_A 2R_B}{L^2-R_A^2-R_B^2}\ll 1
\label{conformal-ratio}
\ee
where $R_{A,B}$ are the radius of the spheres $A,B$ and $L$ is the distance between them. Therefore, we have the freedom to work in a conformal frame where $R_A \sim L\gg R_B$, see Fig. \ref{two-equivalent-frame-setup}, so that the fine-details in the shape-response of $I_{A,B}$ on $A$ are optimally pronounced. A key input into our calculation is the modular Hamiltonian version of the OPE expansion (\ref{eq:MI_OPE}) derived in \cite{Faulkner2021Aug}. This is because at the first order, the shape perturbation theory can be implemented in the context of the entanglement first law, which requires knowledge about the unperturbed modular Hamiltonian.

The paper is organized as follows. In section {\ref{sec:methodology}}, we recall some basic ingredients regarding shape perturbation theory and its application through the entanglement first law for computing linear responses. In section {\ref{sec:modular-Hamiltonian}}, we summarize previous results for mutual information and the corresponding modular Hamiltonians in the OPE limit, which our calculations will be based upon. In section {\ref{sec:main-text}}, we carry out the main computation and obtain results for linear response of mutual information under shape deformation on one of the spheres. During this we identify some subtleties associated with the integral expression of modular Hamiltonian in terms of local operators. In section {\ref{sec:extremality}}, we explore the implications of our results from a symmetry point of view, and discuss the extremality property under shape deformations for mutual information on spheres. In section {\ref{sec:discuss}}, we conclude the paper with some discussions and suggestions for future directions.

\section{Shape perturbation theory and the entanglement first law}\label{sec:methodology}
In this section, we quickly recall the necessary ingredients needed for our computations. More details can be found in \cite{Rosenhaus2014Dec, Rosenhaus2015Feb,Mezei2015Feb, Mezei2015Feb02,Faulkner2016Apr}. 

\subsection{Shape perturbation theory}
\begin{figure}[h]
    \centering
    \includegraphics[scale=0.60]{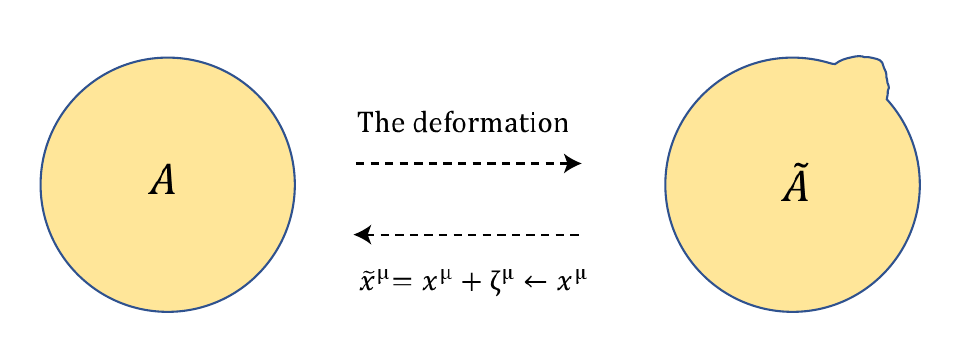}
    \caption{Relating shape deformation to metric deformation via a corresponding diffeomorphism.}
    \label{the deformation of subregion-setup}
\end{figure}

We are only interested in the vacuum states, whose wave-functionals in QFTs we assume can be represented by Euclidean path-integrals in the lower-half-planes. As a result, the reduced density matrix elements on subsystem $A$ can be written in terms of the Euclidean path-integrals with branch-cuts on $A$:
\be
\langle \alpha|\rho_A |\beta\rangle =  \int_{\Phi\left(A^{-}\right)=\beta}^{\Phi\left(A^{+}\right)=\alpha}[\mathcal{D} \Phi] e^{-I_{E}\left(g, \Phi \right)}
\ee
where $\Phi$ denotes the quantum fields collectively. Now we perform an infinitesimal shape deformation of the entangling region $A \rightarrow \tilde{A}$.  For such a deformation, we can always find a corresponding coordinates transformation  $\mathcal{F}:x^\mu\to \tilde{x}^\mu(x)=x^\mu + \zeta^\mu(x)$ on the whole Euclidean space-time such that $\mathcal{F}(\tilde{A})= A$ (see Fig \ref{the deformation of subregion-setup}).

Via diffeomorphism equivalence we can then trade the shape-deformation for a metric deformation $g_{\mu\nu}\to \tilde{g}_{\mu\nu}$ and write the reduced density matrix elements as:
\be
\langle \alpha|\rho_{\tilde{A}}^g |\beta\rangle =\langle \tilde{\alpha} | \rho_A^{\tilde{g}} |\tilde{\beta}\rangle  = \int_{\Phi\left(A^{-}\right)=\tilde{\beta} }^{\Phi\left(A^{+}\right)=\tilde{\alpha}}[\mathcal{D} \Phi] e^{-I_{E}\left(\tilde{g}, \Phi \right)}
\ee
where:
\be
\tilde{\alpha} = \alpha\circ \mathcal{F},\;\;\tilde{\beta} = \beta\circ \mathcal{F},\;\; \tilde{g}_{\mu\nu}=\nabla_\mu \tilde{x}^\alpha \nabla _{\nu} \tilde{x}^\beta g_{\alpha\beta}
\ee
For deformations $\zeta^\mu(x)$ that are small, we can extract the change in the reduced density matrix to leading order in $\zeta$. They are simply given by linear response of the metric deformation, \ie proportional to the stress tensor:
\bea
\delta \rho_A &=& U^\dagger\circ\rho_{\tilde{A}}\circ U -\rho_A = \frac{1}{2}\int \delta g^{\mu\nu} \rho_A\;\hat{T}_{\mu\nu} + \mathcal{O}\left(\zeta^2\right)\nonumber\\
\delta g^{\mu\nu} &=& 2\nabla^\mu\zeta^\nu,\;\;U |\tilde{\alpha} \rangle = |\alpha\rangle
 \label{shape-deformation}
\eea

\subsection{Shape response from the entanglement first law}
In order to compute the change in entanglement entropy, recall that under any change of the reduced density matrix: $\tilde{\rho}_A \to \rho_A + \lambda\; \delta \rho_A,\;\lambda \ll 1$, the first order change is dictated by the so-called the entanglement first law:
\be
\delta S_{A} = \lambda\; \text{tr}\delta \rho_A H_{A} + \mathcal{O}(\lambda^2),\;\;H_A = -\ln{\rho_A}
\ee
This is a consequence of the positivity constraint for the relative entropy between $\rho_A$ and $\tilde{\rho}_A$. Combining the entanglement first law with the shape perturbation theory \eqref{shape-deformation}, one can write the linear response of entanglement entropy under shape deformation as:
\bea
\delta S_A &=& S_{\tilde{A}}-S_A = -\text{tr}\left(U^\dagger \circ \rho_{\tilde{A}} \circ U\right) \ln{\left(U^\dagger \circ\ln{\rho_{\tilde{A}}}\circ U\right)} + \text{tr}\rho_A \ln{\rho_A}\nonumber\\
&=& \text{tr}\; \delta \rho_A\; H_A + \mathcal{O}\left(\zeta^2\right) = \int dx \sqrt{g} \,\  \nabla^\mu \zeta^\nu(x)\; \text{tr} \left\lbrace \rho_A T_{\mu\nu}(x) H_A\right\rbrace + \mathcal{O}(\zeta^2)\nonumber\\
&=& \int dx \sqrt{g}\,\ \nabla^\mu \zeta^\nu(x)\; \left\langle T_{\mu\nu}(x) H_A\right\rangle + \mathcal{O}(\zeta^2)
\label{defomation-int}
\eea
We see that in order to apply \eqref{defomation-int}, we need to have the knowledge of modular Hamiltonians, and compute its correlation functions with the stress tensor. 

\subsection{Warming up: shape dependence of single-interval entanglement entropy in 2d CFT}\label{sec:1-interval}
As a warm-up exercise for subsequent computations, we revisit the shape dependence of single-interval entanglement entropy in 2d CFT using the entanglement first law. The result is of course known, and the computation for single-sphere in arbitrary dimensional CFTs has been done, e.g. in \cite{Mezei2015Feb, Mezei2015Feb02}. The purpose is to recognize patterns and subtleties that may appear in the computations for mutual information.

\begin{figure}[h]
    \centering
    \includegraphics[scale=0.60]{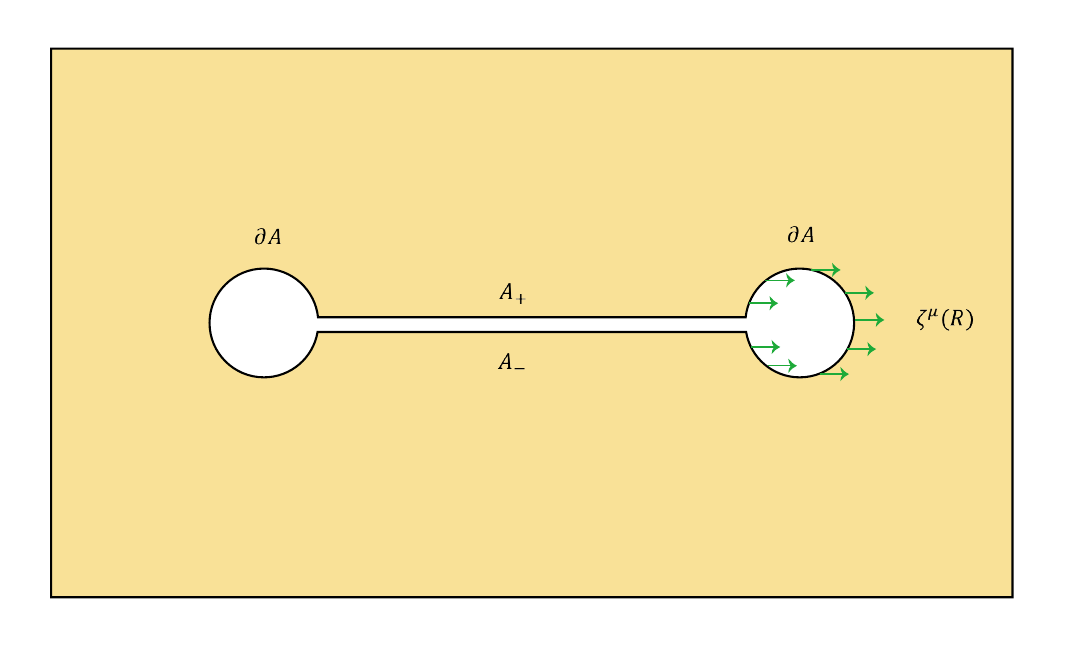}
    \caption{The UV-cut surface at the $\partial{A}$ of the single interval $(-R, R) $ in 2D CFT. For the deformation vector field $\zeta(x)$, we only show its restriction on $\partial A $ which is relevant for our calculations.}
    \label{The replica setup}
\end{figure}
 We label the single-interval subregion $A $ by:
\be
 A \equiv \{ -R \le x \le R\}, \quad \tau= 0
\ee
 From \cite{Casini2011May},  the modular Hamiltonian for $A$ is given by:
\be
H_{A}=2 \pi \int_{A} \frac{R^{2}-x^{2}}{2 R} T_{00}(x) d x  
 \label{single-mH}
\ee 
Now we deform the single-interval subregion $A$ by moving its right end point $R \rightarrow R + \zeta^x(R) $. This can be realized by a deformation vector field $\zeta(x)$ supported about the end point $ R$. The linear response of $S_{A}$ under this deformation can be computed by \eqref{defomation-int}:
\be
 \label{single-interval-delta-S}
 \begin{aligned}
     \delta S_{A} &=\int d^{2} x \,\ \partial^{\mu} \zeta^{\nu}(x)\left\langle T_{\mu \nu}(x) H_{A} \right\rangle \\
     &= \int d^{2} x \,\ \partial^{\mu}\bigg(\zeta^{\nu}(x)\left\langle T_{\mu \nu}(x) H_{A} \right\rangle\bigg) - \int d^{2} x  \,\ \zeta^{\nu}(x) \left\langle  \partial^{\mu}T_{\mu \nu}(x) H_{A} \right\rangle \\
     & = \int d^{2} x \,\ \partial^{\mu}\bigg(\zeta^{\nu}(x)\left\langle T_{\mu \nu}(x) H_{A} \right\rangle \bigg)
     =  \int_{\mathcal{E}} d \bar{x} \sqrt{h} \,\ n^{\mu} \zeta^{\nu}(\bar{x})\left\langle T_{\mu \nu}(\bar{x}) H_{A}\right\rangle \\
 \end{aligned}
\ee
In the second line, we perform an integration by parts, the second term vanishes by Ward-identity up to possible contact terms.  In this case, we can use the local nature of modular flow by $H_{A}$ and infer that:
\begin{equation}
    \begin{aligned}
        \left\langle  \partial^{\mu}T_{\mu \nu}(x) H_{A} \right\rangle &= \left\langle e^{i H_{A} s} \partial^{\mu}T_{\mu \nu}(x) e^{-i H_{A} s} H_{A} \right\rangle 
        \propto  \left\langle  \partial^{\mu}T_{\mu \nu}(x_{s})  H_{A} \right\rangle 
    \end{aligned}    
\end{equation}
where $x_{s}$ is the modular flow trajectory of \eqref{single-mH}. Therefore one can always apply appropriate modular flow to move $\partial^{\mu} T_{\mu \nu}$ away from the support of $H_{A}$, and conclude that there will be no contact term contribution. We can therefore drop this term.  As a total derivative, the first term is reduced to a boundary integral by Stokes theorem. The boundary is imposed as a UV regulator for the calculation of entanglement entropy. In this case, the boundary $\mathcal{E}$  is a codimension one tube of radius $\epsilon$  enclosing the entangling surface $\partial A $, \ie $\mathcal{E} = \partial A \times S^{1}$. For our case, the entangling surface $ \partial A $ is just the right end point. Now we proceed by plugging in the explicit form of $H_{A}$ into \eqref{single-interval-delta-S},  then the boundary integral becomes the contour integral in $z = x + i\tau $:
\be
\begin{aligned}
	\delta S_{A} &=  i \oint_{z=R} \frac{d z}{2 \pi} \;  \zeta^{z}(z) \left\langle  T(z) H_{A} \right\rangle + h.c. \\
	 &=  i \zeta^{z}(R) \oint_{z=R} \frac{d z}{2 \pi} \int_{-R}^{ R} d w \;\left( \frac{R^{2}-w^{2}}{2 R}  \right) \langle T(z) T(w)\rangle + h.c.  \\
     & = i \zeta^{z}(R) \oint_{z=R} \frac{d z}{2 \pi} \int_{-R}^{ R} d w \; \left( \frac{R^{2}-w^{2}}{2 R}  \right) \frac{c/2}{(z-w)^{4}} + h.c.  \\
     & = i \zeta^{z}(R) \oint_{z=R} \frac{d z}{2 \pi} \; \left[\frac{c R^{2}}{3(R-z)^{2}(R+z)^{2}} \right]  + h.c. \\
	& = \frac{c}{ 6 R} \left( \zeta^{z}(R) + \zeta^{\bar{z}}(R)\right) = \frac{c}{ 6 R} \zeta^x(R)
\end{aligned}
 \label{single-interval-deform}
\ee
which agrees with the well-known result:
\begin{equation}
    S_{A}=\frac{c}{3} \log \left(\frac{2 R}{\epsilon}\right)
\end{equation}
In the second line of \eqref{single-interval-deform} , we have replaced $\zeta^{z}(z)$ by its  average $\zeta^{z}(R)$ because we are only interested in the mode that translates the tube $\mathcal{E}$, instead of changing the geometry of the tube which the nonzero modes of $\zeta$ effectuate. 

 We summarize the take-away message of \eqref{single-interval-deform} as follows. In the context of the entanglement first law, the linear response of entanglement entropy is proportional to a contour integral of stress tensor correlator with modular Hamiltonian:
 \begin{equation}
    \delta S_{A} \propto \oint_{z \sim \partial A} {d z} \; \langle T(z) H_{A}\rangle
 \end{equation}
 This only picks up the residue of simple pole in $\langle T(z) H_{A}\rangle$, schematically of the form:
  \begin{equation}
     \langle T(z) H_{A}\rangle = \cdots + \frac{Res}{z-\partial A} + \cdots
  \end{equation}
In cases where the modular Hamiltonian  can be written as  integrals of local operators like \eqref{single-mH}, such poles emerge only after completing all the integrals in the modular Hamiltonian. For example, switching the order of $ \int {d w} $ and $\oint_{z=\partial A} {d z}$  in \eqref{single-interval-deform} gives zero instead. This point is important later when we study shape dependence of mutual information using the entanglement first law.
 
\section{ Mutual information and modular Hamiltonian in the OPE limit}\label{sec:modular-Hamiltonian}
    Our goal is to study  the shape dependence of mutual information perturbatively about two spheres in the CFT vacuum. Similar to the computation of  $\delta S_{A}$ in Sec \eqref{sec:1-interval}, this can be done via the entanglement first law. The pre-requisite for such an approach is explicit knowledge of the unperturbed result for both  the mutual information and its modular Hamiltonians. However,  mutual information  in general is highly theory dependent, even for the case of our interest, \ie two spheres in the vacuum.  
    Explicit results are only accessible in the so-called OPE limit, in which the distance between  two spheres are much larger than their radii. As a result, in this paper we shall also focus on the shape dependence of mutual information in the same (or conformally equivalent) limit (see Fig \ref{two-equivalent-frame-setup}). In this section, we review the results in \cite{Faulkner2016Aug,Faulkner2021Aug}  calculations of  for the OPE limit of  mutual information and the corresponding  modular Hamiltonian using the replica technique.
 
   We  consider two disjoint spherical subregions $A$ and $B$ with radius $R_{A}, R_{B}$  separated by a distance $L$ (see Fig \ref{two-equivalent-frame-setup}). The mutual information $I_{A,B}$ can be obtained by the replica trick as: $ I_{A, B} \equiv \lim _{n \rightarrow 1} I_{n}(A, B) $, where:
   \begin{equation}
   	I_{n}(A, B) \equiv S_{n}(A)+S_{n}(B)-S_{n}(A B)=\frac{1}{1-n} \log \left(\frac{\operatorname{tr} \rho_{A}^{n} \operatorname{tr} \rho_{B}^{n}}{\operatorname{tr} \rho_{A B}^{n}}\right)
   \end{equation}
   is the R{\'e}nyi mutual information.
  For QFTs defined on Euclidean spacetime $\mathcal{M}$, we can rewrite these traces as partition functions on appropriately constructed  conifolds, \ie
  \begin{equation}
  	I_{n}(A, B)=\frac{1}{n-1} \log \left(\frac{Z\left(\mathcal{C}_{n}^{A \cup B}\right) Z^{n}(\mathcal{M})}{Z\left(\mathcal{C}_{n}^{A }\right) Z\left(\mathcal{C}_{n}^{B}\right)}\right)
  	\label{renyi-MI}
  \end{equation}
   where  $ \mathcal{C}_{n}^{X} $  is constructed from $n$-copies of $\mathcal{M}$,  glued across subregion $ X$.  Equivalently, we can write the partition functions 
  on conifold as expectation values of twist operators in the corresponding orbifold theory: $CFT^{\otimes n}/\mathds{Z}_{n}$,  defined on $\mathcal{M}^{n}$:
   \begin{equation}
   	  \begin{aligned}
   	  	& Z\left(\mathcal{C}_{n}^{A }\right) =  Z^{n} \left\langle \Sigma_{n}^{A} \right\rangle_{\mathcal{M}^{n}},\quad  Z\left(\mathcal{C}_{n}^{B }\right) =  Z^{n} \left\langle \Sigma_{n}^{B} \right\rangle_{\mathcal{M}^{n}} \\
   	  	&  Z\left(\mathcal{C}_{n}^{A \cup B }\right) =  Z^{n} \left\langle \Sigma_{n}^{A} \Sigma_{n}^{B} \right\rangle_{\mathcal{M}^{n}} .
   	  \end{aligned}
   \end{equation}

     Analogous to ordinary correlation functions, the twist operators in $\left\langle \Sigma_{n}^{A} \Sigma_{n}^{B} \right\rangle_{\mathcal{M}^{n}} $ also admit operator product expansions into local operators in $CFT^{\otimes n}/\mathds{Z}_{n}$ :  
    \begin{equation}\label{eq:twist_OPE}
    	\begin{aligned}
    	   \Sigma_{n}^{A,B} &= \frac{Z\left(\mathcal{C}_{n}^{A,B}\right)}{Z^{n}} \bigg( 1 + \sum_{\left\{k_{j}\right\}} C_{\left\{k_{j}\right\}}^{A,B} \prod_{j=0}^{n-1} \Phi_{k_{j}}^{(j)}\left(r_{A,B}\right) \bigg) 
    	\end{aligned}
    \end{equation}
 \ie expanding $\Sigma_{n}^{A,B}$ into a complete set of local operators at $r_{A,B}$, where in \eqref{eq:twist_OPE} we have written the leading identity operator separately. Assuming a  $\mathds{Z}_{2}$ symmetry for the underlying CFT, the coefficients for single copy operators vanish. As a result, the first few orders in the OPE take the form :
 
 \begin{equation}
     \Sigma_{n}^{A,B} = \left\langle \Sigma_{n}^{A,B} \right\rangle_{\mathcal{M}^{n}} \bigg( 1+ \frac{1}{2}( 2 R_{A,B})^{2 \Delta} \sum_{j \neq k}^{n-1} c_{j-k} \mathcal{O}^{(j)} \mathcal{O}^{(k)}+ \cdots \bigg)
 \end{equation}
 where $\mathcal{O}$ is a primary operator with the lowest scaling dimension $\Delta$, and it enters the OPE by appearing twice in different replica $j \neq k$. We call such form of  OPE the bi-local channel. Due to replica symmetry, the OPE coefficients only depend on $j-k$. The subsequent $\cdots$ denotes contributions from other operators  in the  bi-local channel or  from multi-local channels. For future convenience, we have rescaled the OPE coefficients with a factor $(2R_{A,B})^{-2\Delta}$  relative to \eqref{eq:twist_OPE}, \ie $ c_{j-k}  = (2 R_{A, B} )^{-2 \Delta} C_{j-k} $. In this paper, we focus on the leading-order contribution to $I_{A,B}$. This corresponds to $\mathcal{O}^{(j)}\mathcal{O}^{(k)}$ composed by the lowest-dimension primary operator $\mathcal{O}$. For simplicity, we begin by assuming $\mathcal{O}$ as a scalar, and will discuss the cases for spinning operators later. 

Analogous to the usual OPE, we can extract the coefficients $c_{j-k}$ by inserting corresponding operators at $r\to\infty$ and inverting (\ref{eq:twist_OPE}):
\bea 
c_{j-k} &=& \lim_{r\to\infty}\left\langle \Sigma^{A,B}_n \mathcal{O}^{(j)}(r)\mathcal{O}^{k}(r)\right\rangle G^{-1}(r)G^{-1}(r) \nonumber\\
&=& \lim_{r\to\infty} \Omega_j^{\Delta}\;\Omega_k^{\Delta}\;G^{-1}(r)\;G^{-1}(r)\;\left\langle \mathcal{O}\left(Z_j\right)\mathcal{O}\left(Z_k\right)\right\rangle_{\mathds{S}_{(2\pi n)}\times\mathds{H}_{d-1}}\nonumber\\
&=& \left\langle \mathcal{O}(\tau_j,0)\mathcal{O}(\tau_k,0)\right\rangle_{\mathds{S}_{(2\pi n)}\times\mathds{H}_{d-1}}\nonumber\\
&=& G_{n}(\tau_i-\tau_j),\;\;\tau_i=(2i+1)\pi
\eea  
The correlation function in the presence of $\Sigma^{A, B}_n$ has been written in terms of thermal correlator $G_n$ in hyperbolic space $\mathds{H}_{d-1}$ at inverse temperature $T^{-1} = 2\pi n$. This is achieved via the coordinate transformation into hyperbolic space coordinates $Z =(\tau,Z^i)\in \mathds{S}_{(2\pi n)}\times \mathds{H}_{d-1}$: 
\bea\label{eq:hyperbolic}
Z^0 &=& \tau = \cos^{-1}\left(\frac{R^2-r^2}{\sqrt{(R^2-r^2)^2+4R^2 r_0^2}}\right)\nonumber\\
Z^i &=& \frac{2R r_i}{\sqrt{(R^2-r^2)^2+4R^2r_0^2}},\;\;i\geq 1
\eea 
From this form, one can easily perform the analytic continuation in $n$ for the replica trick and obtain the contribution to the mutual information: 
\bea \label{MI-Rindler}
&& I_{A,B} = \lim_{n\to 1} \frac{1}{1-n} \bigg(\sum_{j\neq k}c_{j-k}^2 \bigg)  \left(\frac{2R_A 2R_B}{L^2} \right)^{2\Delta}\nonumber\\
&=& \left(\frac{2R_A 2R_B}{L^2} \right)^{2\Delta} \int^\infty_{-\infty} ds\frac{G_1(-is+\pi)^2}{4\cosh^{2}\left(s/2\right)}\nonumber \\ 
& =& \left(\frac{2R_A 2R_B}{L^2} \right)^{2\Delta} \frac{\sqrt{\pi}}{4^{2\Delta + 1}} \frac{\Gamma(2\Delta+1)}{\Gamma(2\Delta + 3/2)} .
\eea 
 The prefactor is just the conformal ratio $\rho$ in \eqref{conformal-ratio} in the OPE limit.  More details of the computations can be found in \cite{Cardy2013, Faulkner2016Aug, Faulkner2021Aug}.
    
  For the application of the entanglement first law, we also need the explicit form  of the modular Hamiltonians in the combination $ \Delta H_{A,B} \equiv H_{A,B}- H_{A} -H_{B} $. 
   In the OPE limit, it admits an expansion similar to that of the mutual information. In this case the OPE is only taken for one of the twist operators $\Sigma_n^B$. After analytic continuation, the local operators from OPE are modular flown by $H_{A}$. The details can be found in \cite{Faulkner2021Aug}. In the end, one obtains the OPE for $\Delta H_{A,B}$ in the bi-local channel of primary scalar operator $\mathcal{O}$ as: 
\bea \label{mH-two-sphere}
    \Delta H_{A,B} &=& -{(2R_{B} )^{2 \Delta}}  \int_{-\infty}^{\infty} {d s} \,\  k_{1}(s) \; \mathcal{O}_{A}(-is) \; \mathcal{O}_{B} \nonumber\\
    &+& \frac{i(2R_{B} )^{2 \Delta}
       		}{2 \pi }
       	 \int_{-\infty}^{\infty} {d s_{j} d s_{k}} \,\ k_{2}(s_{j}, s_{k}) \; \mathcal{O}_{A}(-is_{k} - i s_{j}) \; \mathcal{O}_{A}(-is_{j}) + \delta H_{BB}\nonumber\\
    k_{1}(s) &=& \frac{1}{4\cosh^2(s/2)} c_1(-is+\pi)\nonumber\\
    k_{2}(s_{j}, s_{k})& =& \frac{1}{4 \cosh^{2} (s_{j}/2)} \left( \frac{1}{e^{s_{k}+ i \epsilon}-1} + \frac{1}{e^{s_{k}+s_{j}}+1} \right) c_{1}(-is_{k} + \epsilon) 
\eea
where $c_1(is)$ is the analytic continuation of $c_{n}$ at $n=1$:
\be
c_{1}(is) =  \left(-4\sinh^2(s/2)\right)^{-\Delta}
\ee 
\begin{figure}[h]
    \centering
    \includegraphics[scale=0.60]{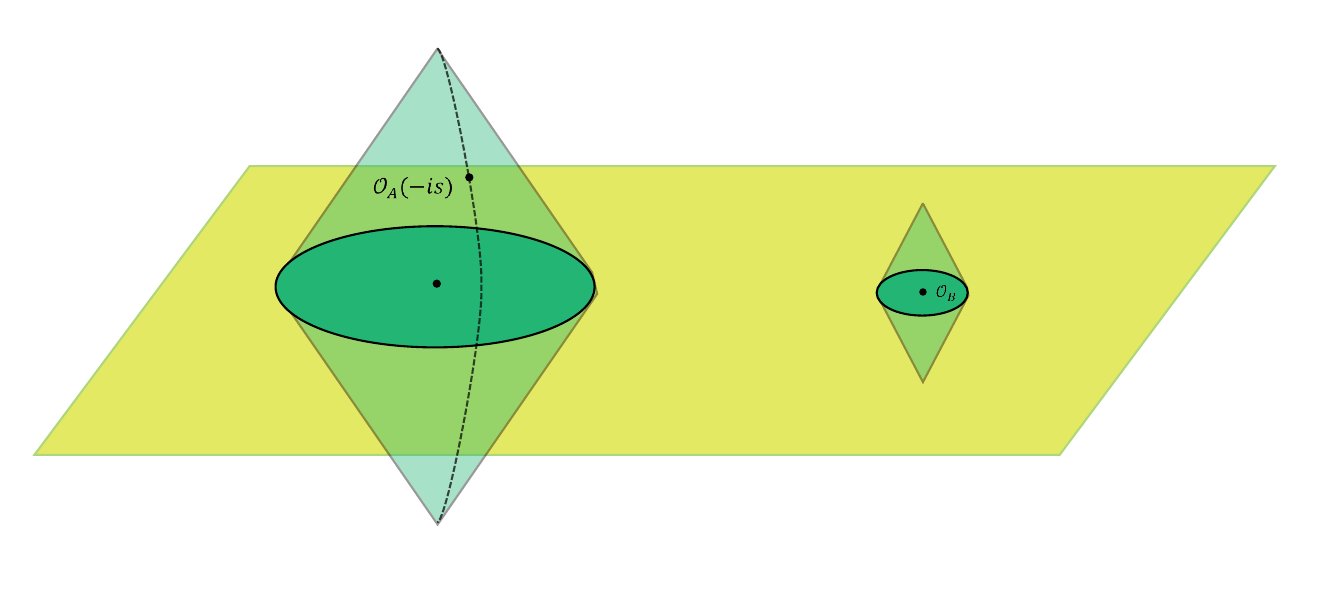}
    \caption{The locations for $\mathcal{O}_{A}(-i s)$ and $\mathcal{O}_{B}$. In the limit of large separation, the modular trajectory of $\mathcal{O}_A(-is)$ almost passes through the center of $A$. }
    \label{double-cone}
\end{figure}

The operator $ \mathcal{O}_{A}(-is) $ is obtained from $\mathcal{O}_{B} = \mathcal{O}(r_{B})$ by first applying a modular reflection in $H_{A}$, followed with a modular flow by real $s$ (see Fig \ref{double-cone}):   $$ \mathcal{O}_{A}(-is)   =\rho_{A}^{is-1 / 2}\; \mathcal{O}_{B} \; \rho_{A}^{-is+1 / 2} .$$
   
   As a check, one can compute the expectation value of the modular Hamiltonian  \eqref{mH-two-sphere} in the vacuum state:

   \begin{equation}
   	\begin{aligned}
   		 \langle \Delta H_{A, B}  \rangle = \langle \Delta H_{A, B}^{(1)}  \rangle + \langle \Delta H_{A, B}^{(2)}  \rangle
   	\end{aligned}
    \label{MI-MH-form}
   \end{equation}
  where $ \langle \Delta H_{A, B}^{(1)}  \rangle $ and $ \langle \Delta H_{A, B}^{(1)}  \rangle $ represents the contribution from the single integral and double integral of Eq \eqref{mH-two-sphere} respectively. It can be found that:
  \begin{equation}
  	\begin{aligned}
  	\langle \Delta H_{A, B}^{(1)}  \rangle  = -2 \langle \Delta H_{A, B}^{(2)}  \rangle 
  	= - 2 \rho^{2 \Delta}  \frac{\sqrt{\pi}}{4^{2\Delta + 1}} \frac{\Gamma(2\Delta+1)}{\Gamma(2\Delta + 3/2)} = -2I_{A,B}.
  	\end{aligned}
  \end{equation}
  which once combined in \eqref{MI-MH-form} agrees with the result for mutual information \eqref{MI-Rindler}.

\section{Shape dependence of mutual information in the OPE limit}\label{sec:main-text}
\begin{figure}[h]
      \centering
      \includegraphics[scale=0.65]{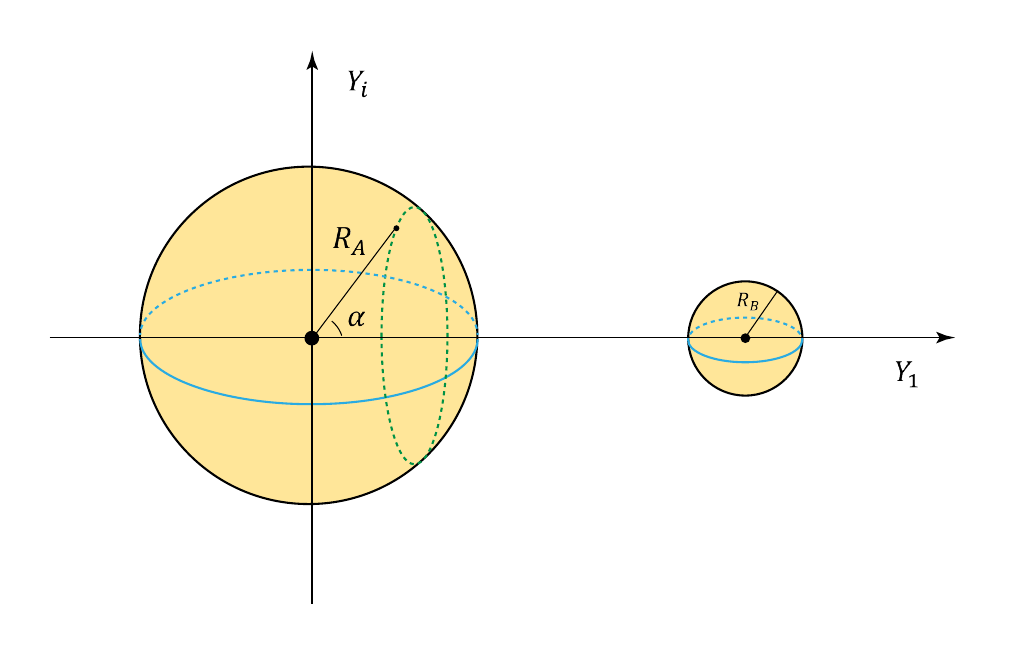}
      \caption{The setup of spherical frame coordinates $Y$. The center of the sphere $A$ has been set to the origin. While that of sphere $B$ is located at $Y_{1}=L, Y_i =0$. The angular coordinate $\alpha$ on $\partial A$ is defined as the angle relative to the $Y_1$ axis.}
      \label{sphere-frame}
  \end{figure}

  In this section, we give the main calculations of this paper. The task is to compute the linear response of mutual information $I_{A,B}$ between two spheres $A$ and $B$ under shape deformation on one of the spheres $A\to \tilde{A}$. The spheres have radius $R_A$ and $R_B$ whose centers are separated by a distance $ L$. We focus on a conformally equivalent version of the OPE limit $L\sim R_{A} \gg R_{B}$ (see Figure \ref{sphere-frame}). 
  
  More specifically, the deformed sphere $\tilde{A}$ can be parametrized by an angular-dependent modification of the radius: $R(\Omega) = R_{A} + \zeta(\Omega) $. In this parametrization, we can work with the ansatz in which the deformation vector field $\zeta^\mu$ that enters the entanglement first law \eqref{defomation-int}  only has a radial component in polar coordinates about the center of $A: \zeta^\mu \propto \delta^\mu_r$. 

Now, let us calculate the shape deformation of mutual information by using the ingredients that we have prepared in previous. Similar to the warm-up exercise in \ref{sec:1-interval}, application of the entanglement first law yields that:
\begin{equation}
    \label{formal-deformation}
    \begin{aligned}
        \delta I_{A,B}  &= \int_{\mathcal{M}} d^{d}Y  \,\ \partial^{\mu} \zeta^{\nu}(Y)\left\langle T_{\mu \nu}(Y) \Delta H_{A, B} \right\rangle \\
        &= \int_{\mathcal{M}} d^{d}Y \,\ \partial^{\mu}\bigg( \zeta^{\nu}(Y)\left\langle T_{\mu \nu}(Y) \Delta H_{A, B} \right\rangle \bigg) - \int_{\mathcal{M}} d^{d}Y \,\ \zeta^{\nu}(Y) \left\langle \partial^{\mu} T_{\mu\nu}(Y) \Delta H_{A,B} \right\rangle \\
        & = \int_{\partial \mathcal{M}} \epsilon \; {d\theta} \;{d}^{d-2} \Omega  \sqrt{h(\Omega)} \,\ n^{\mu} \zeta^{\nu}  \langle  T_{\mu \nu}(\theta,\Omega) \Delta H_{A, B} \rangle , 
    \end{aligned}
\end{equation}
where $\partial \mathcal{M} \sim S^{1} \times S^{d-2} $ is the UV cut-off surface in the form of codimension one tube of radius $\epsilon$ enclosing the spherical boundary $\partial A $, and $(\theta, \Omega_{d-2}) $ are coordinates on $\partial \mathcal{M}$. Similar to \eqref{single-interval-delta-S}, here we can also drop the contact term in the second line. The reason is that  the OPE of  $\Delta H_{A,B}$ consists of bi-local operators whose location we can avoid by restricting the support of $\zeta^{\mu}$ in the bulk.

 From Eq.  \eqref{mH-two-sphere}, we see that the OPE limit of modular Hamiltonian $ \Delta H_{A, B} $ is written as the integral of local operators which are flowed by the modular Hamiltonian of spherical subregion $ A $.  In terms of actual calculations, this is not convenient because the expressions for the modular trajectories are very complicated.  We can simplify the calculations by exploiting the conformal symmetry. More specifically, we observe that \eqref{formal-deformation} involves integrals of CFT correlators of the form  
 \begin{equation}
    \langle T_{\mu\nu} \mathcal{O}(-is_{1})\mathcal{O}(-is_{2}) \rangle =  \langle T_{\mu\nu}\rho_{A}^{is} \mathcal{O}(Y_{B}) \rho_{A}^{-is_{1}+ is_{2}}\mathcal{O}(Y_{B}) \rho_{A}^{-is_{2}} \rangle
    \label{three-pt-correlation}
 \end{equation}
where $Y_{B} $ is the center of a small sphere $B$. For CFTs, there is a conformal transformation $\mathcal{U}$ that maps from $A$ to the Rindler half plane. The modular evolution operator can be related to  Rindler boost  $K$ by: 
\begin{equation}
    \label{rho-to-rho}
    \rho_{A}^{is} = \mathcal{U}^{-1} e^{is K} \mathcal{U} 
\end{equation}
Plugging \eqref{rho-to-rho} into \eqref{three-pt-correlation}, and using the invariance of the conformal vacuum under $\mathcal{U}$, we can rewrite the correlator in the Rindler frame:
 \begin{equation}
     \begin{aligned}
          \langle T_{\mu\nu} \mathcal{O}(-is_{1})\mathcal{O}(-is_{2}) \rangle &=  \langle T_{\mu\nu} \mathcal{U}^{-1} e^{is_{1} K} \mathcal{U}  \mathcal{O} \mathcal{U}^{-1} e^{-is_{1} K + i s_{2}K} \mathcal{U} \mathcal{O} \mathcal{U}^{-1} e^{-is_{2} K} \mathcal{U}  \rangle \\
          &= \langle \mathcal{U}  T_{\mu\nu} \mathcal{U}^{-1} e^{is_{1} K} \mathcal{U}  \mathcal{O} \mathcal{U}^{-1} e^{-is_{1} K + i s_{2}K} \mathcal{U} \mathcal{O} \mathcal{U}^{-1} e^{-is_{2} K} \rangle \\
          &= \langle   \tilde{T}_{\mu\nu}  e^{is_{1} K} \tilde{\mathcal{O}}  e^{-is_{1} K + i s_{2}K} \tilde{\mathcal{O} } e^{-is_{2} K} \rangle \\
           &= \langle   \tilde{T}_{\mu\nu}  \tilde{\mathcal{O}}(-is_{1})  \tilde{\mathcal{O}}(-is_{2})  \rangle \\
     \end{aligned}    
     \label{three-pt-correlation-Rindler}
 \end{equation}
 where $\tilde{T}_{\mu\nu} =  \mathcal{U}^{-1} \; T_{\mu\nu} \; \mathcal{U}$,  and $\tilde{\mathcal{O}}(-is) = \Lambda(Y_{B})^{-\Delta} \; e^{iK s} \mathcal{O}\Big(\mathcal{U}(Y_{B})\Big) e^{-iKs}$ are boosted  operators in Rindler frame rescaled by the conformal factor  $\Lambda(Y_{B})$ of $\mathcal{U}$.
  
 \subsection{Mapping spherical frame to Rindler frame}
  The conformal transformation $\mathcal{U}$ from the spherical frame coordinates $Y$ to the Rindler frame coordinate $X$  is given explicitly by the following form:
    \begin{equation}
      \begin{aligned}
         C_{I} X^{\mu} &=  \frac{\tilde{Y}^{\mu} + \big(\tilde{Y} \cdot \tilde{Y}\big) C^{\mu}}{1+2\big(C \cdot \tilde{Y}\big)+ \big(C\cdot C \big) \big(\tilde{Y} \cdot \tilde{Y}\big) }, \\ 
          \tilde{Y}^{\mu} &= Y^{\mu}  - R_{A} \delta^{\mu}_{1}, \quad C^{\mu} = {\delta_{1}^{\mu}}/{2 R_{A}} \\
            C_{I} &=  \frac{2 R_{A} \big( L^{2} - R_{A}^{2} - R_{B}^{2}\big)}{\big(L+ R_{A} + R_{B} \big) \big(L+ R_{A} - R_{B} \big)}
      \end{aligned}
      \label{tranf-ball-halfspace}
    \end{equation} 
   The map is arranged such that  under $\mathcal{U}$ the spherical subregion $ A $ becomes the left half space region  $  x^{1}<0  $
  and  the spherical subregion $ B $ becomes a new sphere with new center at the canonical position $X_{B} = (0,1,0,\cdots, 0)$ and new radius is given by (see Fig \ref{Rindler-frame}):
  \begin{figure}[h]
      \centering
      \includegraphics[scale=0.55]{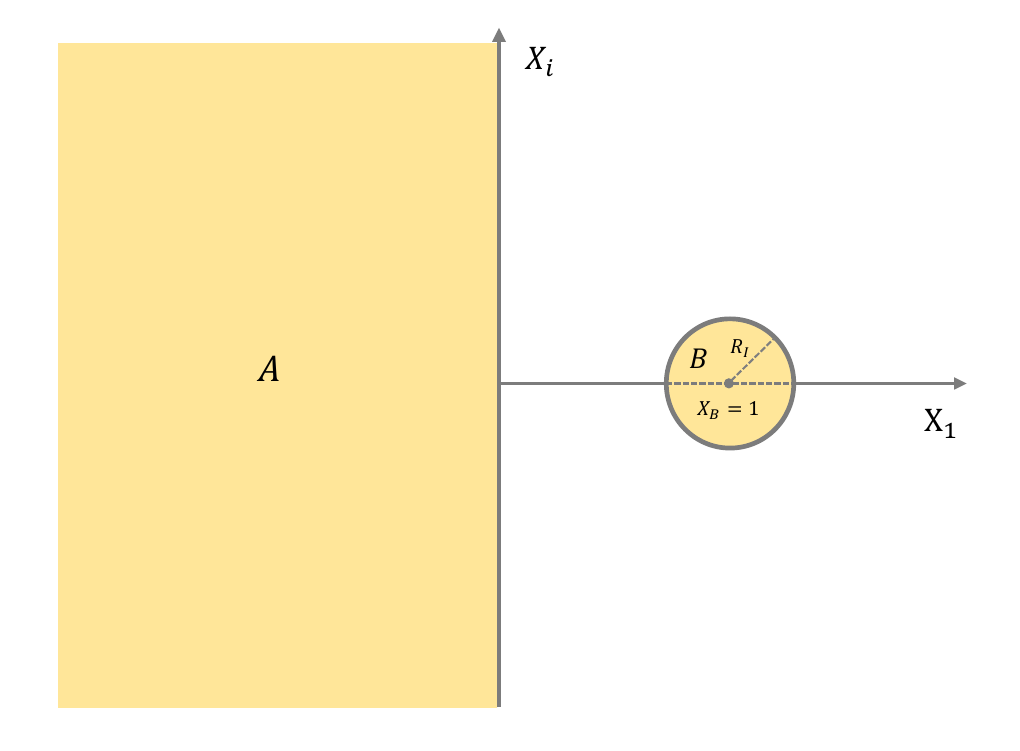}
      \caption{Rindler frame: subregion $A$ is the half space $X_1<0$; subregion $B$ is a sphere whose center is located at $ X_{B} = (0,1,0,...,0)$ with small radius $R_I\ll 1$. }
      \label{Rindler-frame}
  \end{figure}
    \begin{equation}
      \begin{aligned}
          R_{I}  = \frac{ 2 R_{A} R_{B}}{\left(L^{2}- R_{A}^{2} -R_{B}^{2}\right)} = \frac{\rho}{2}
      \end{aligned}
      \label{radius-1}
    \end{equation}
   The corresponding conformal factor is given by: $$  \Lambda(Y)  = C_{I} \left(1+2(C \cdot \tilde{Y})+C^{2} \tilde{Y}^{2} \right) $$  
   To utilize \eqref{three-pt-correlation-Rindler}, we need to keep track of  how the UV cut-off surface transform across $\mathcal{U}$. To proceed, we parametrize the cut-off tube in spherical frame $Y^{\mu}$ (see Fig \ref{sphere-frame}) as: 
  \begin{equation}
      \begin{aligned}
          & Y^{0} = \epsilon \sin \theta,	\qquad Y_{1} = \Big( R_{A} + \epsilon \cos \theta \Big) \cos\alpha \\
          & Y_{i} = \Big( R_{A} + \epsilon \cos \theta \Big) (\sin\alpha ) e_{i}, 	\qquad  \sum_{i \ge 2} e_{i}^{2} = 1 ,
      \end{aligned}
  \end{equation}
  where $\epsilon$ is the UV cut-off scale. Under the  transformation of \eqref{tranf-ball-halfspace}, 
   the UV cut-off manifold $   S^{1} \times S^{d-2} $ is mapped to  a tube $S^{1} \times \mathds{R}^{d-2}$ by the following relation:
   \begin{equation}
       \begin{aligned}
           & C_{I} X^{0}=\epsilon \frac{2 \sin \theta}{1+\cos \alpha}+\mathcal{O}\left(\epsilon^{2}\right) \\
           &C_{I} X_{1}=\epsilon \frac{2 \cos \theta}{1+\cos \alpha}+\mathcal{O}\left(\epsilon^{2}\right) \\
           &C_{I} X_{i}=\frac{2 R_{A} \sin \alpha}{1+\cos \alpha} e_{i}+\mathcal{O}\left(\epsilon^{2}\right), \\
           &\sum e_{i}^{2}=1
       \end{aligned}
      \label{cooridinate-relation}
   \end{equation}
  The geometry of the mapped tube in Rindler frame  is changed such that it has  $X_{i}$-dependent radius (see Fig \ref{UV-cut-off-tube}): 
    \begin{figure}[h]
       \centering
       \includegraphics[scale=0.42]{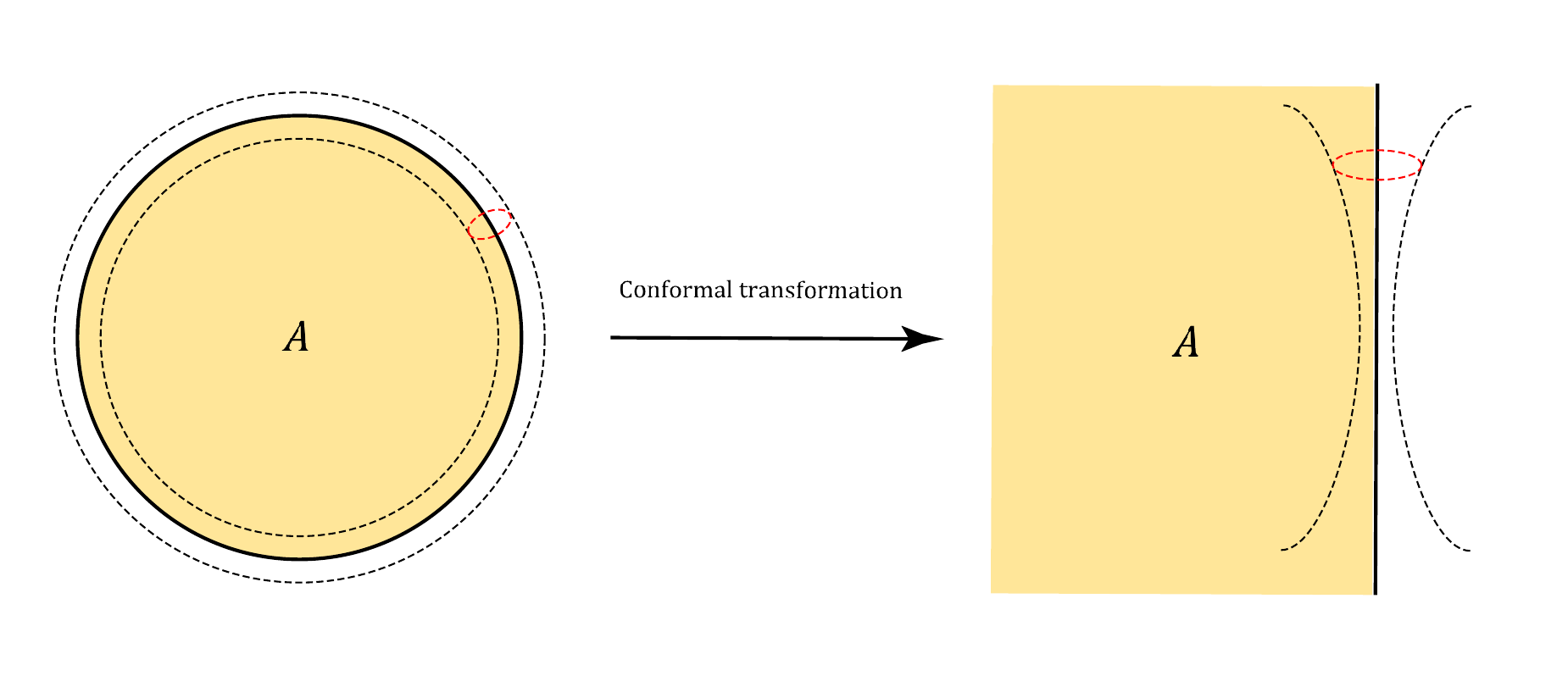}
       \caption{The UV cut-off tube in two conformally equivalent frames. Left: the UV cut-off tube of constant radius in the original spherical frame. Right: the mapped UV cut-off tube of ``trumpet" shape in the Rindler frame.}
       \label{UV-cut-off-tube}
   \end{figure}
   \begin{equation}
      \tilde{\epsilon}(X_{i})  = \epsilon \Big(1+ \frac{C_{I}^{2}}{4R_{A}^{2}} \sum_{i \ge 2} X_{i}^{2} \Big)   {C_{I}}^{-1}
      \label{new-UV-cut-off}
   \end{equation}
    where $(0, 0, X_{i})$ is a corresponding point on the Rindler plane $\mathds{R}^{d-2}$ .  
    
    Now, we can plug the explicit form of $\mathcal{U}$ into \eqref{three-pt-correlation-Rindler}, the correlator $ \langle T_{\mu\nu} \mathcal{O}(-is_{1})\mathcal{O}(-is_{2}) \rangle $ can be explicit written as boosted correlators in Rindler frame:
     \begin{equation*}
        \begin{aligned}
            & n^{\mu} \zeta^{\nu}(\theta,\Omega)  \langle T_{\mu \nu}(\theta,\Omega)  \mathcal{O}(s_{1}) \mathcal{O}(s_{2}) \rangle \\
            & =  {\Lambda(Y_{B})}^{-2\Delta} {\Lambda(\theta,\Omega)}^{-(d-2)}  \left(\frac{d X^{\alpha}}{d Y^{\mu}} \frac{d X^{\beta}}{d Y^{\nu}}\right) \Big|_{\theta,\Omega } n^{\mu} \zeta^{\nu}(\theta,\Omega) \left\langle  T_{\alpha \beta}({\theta,\Omega})  \mathcal{O}(X[s_{1}]) \mathcal{O}(X[s_{2}] ) \right\rangle
        \end{aligned}
    \end{equation*}  
     where 
     \begin{equation}
          X[s]^{\pm} = e^{\pm s},\quad X^{i} = 0
     \end{equation}
    are the boosted coordinates, and the conformal factors are given by:
    \begin{equation}
        \begin{aligned}
             &\Lambda(Y_{B}) =  \frac{ C_{I}\left(L+R_{A}\right)^{2}}{4 R_{A}^{2}} \\
            & \Lambda(\theta,\Omega) =  C_{I} \left(1+  C_{I}^{2} \frac{\sum_{i \geq 2}X_{i}^{2}(\theta,\Omega)}{4 R_{A}^{2}}\right)^{-1}
        \end{aligned}
    \end{equation}
     The factor $\Lambda(Y_{B})^{-2\Delta}$ will combine with the $(2R_{B})^{2\Delta}$ in the OPE coefficient to form the conformal ratio $\rho^{2\Delta}$ in the limit $R_{B} \to 0$; the factor  
   $ \Lambda(\theta,\Omega) $ reflects the geometry new cut-off tube encoded in \eqref{new-UV-cut-off}.
    Gather all pieces together, we can rewrite $\delta I_{A,B}$  as follows: 
    \begin{equation}
        \begin{aligned}
            \delta I_{A,B}
            &=  \int  \tilde{\epsilon} {d}\theta \int {d}^{d-2}X_{i} \,\   \tilde{n}^{\mu} \xi^{\nu} \left\langle T_{\mu\nu }(\theta,\Omega) \Delta \tilde{H}_{A, B} \right\rangle \\
        \end{aligned}
        \label{main-relation}
    \end{equation}
  where $ \tilde{n}^{\mu} =   -e^{i\theta} \partial_{z} -e^{-i\theta} \partial_{\bar{z}}$, $z =  X^{1} + iX^{0}, \,\ \bar{z} =  X^{1} -i X^{0} $ is just the unit normal vector for new UV cut-off tube, and  
  \begin{equation}
      \xi^{\mu}(X) = \xi(X) \left( \partial_{z} + \partial_{\bar{z}}\right), \quad \xi(X\small) =   \Big( 1 +  \frac{C_{I}^{2}}{4R_{A}^{2}} \sum_{i \ge 2} X_{i}^{2}(\theta,\Omega) \Big)   {C_{I}}^{-1}  {\zeta}(\theta,\Omega) 
      \label{relation-of-deformation-vector}
  \end{equation}
  is the ``effective" deformation vector field that appears in the Rindler frame. Furthermore, $\Delta \tilde{H}_{A, B}$  is now in terms of boosted operators in Rindler frame:
\begin{equation}
    \begin{aligned}
        \Delta \tilde{H}_{A, B}=&-\rho^{2 \Delta} \int_{-\infty}^{\infty} d s \;  k_{1}(s) \;  \mathcal{O}\left(-X[s]\right) \mathcal{O}\left(X[0]\right) \\
        &+\frac{i\rho^{2 \Delta}}{2 \pi} \int_{-\infty}^{\infty} d s_{j} \; d s_{k} \; k_{2}\left(s_{k}, s_{j}\right) \;  \mathcal{O}\left(-X[s_{j}+s_{k}]\right) \mathcal{O}\left(-X[s_{j}]\right)
    \end{aligned}
\end{equation}
 Therefore, all computations can be carried out in the Rindler frame, we simply need to keep track of the mapping between conformally related deformation vector fields \eqref{relation-of-deformation-vector}.

\subsection{Shape dependence in Rindler frame}
  Now, we compute the shape dependence of mutual information \eqref{main-relation} by working in the Rindler frame, where the modular flow becomes simply the boost about the half plane $X_{1} < 0 $. Similar to the warm-up exercise, we can write the integral on the cut-off tube in terms of contour integrals in  $z =  X^{1} + iX^{0}, \,\ \bar{z} =  X^{1} -i X^{0} $ :
  \begin{equation}
      \begin{aligned}
          \delta I_{A, B}&= i \int {d}^{d-2} X^{i} \,\ \xi(X^{i})\oint_{|z|=\tilde{\epsilon}} {d}z \,\  \left\langle  \bigg(T_{zz}(z,\bar{z}, X^{i})  + T_{z\bar{z}}(z,\bar{z}, X^{i})\bigg) \Delta \tilde{H}_{A, B} \right\rangle + h.c. \\
      \end{aligned}
   \label{Rindle_I}
  \end{equation}
Notice that in writing \eqref{Rindle_I}, we have taken the deformation vector field $\xi(X^{i})$ outside the contour integral $\oint {d z}$ by the zero-mode substitution: $$\xi(z,\bar{z}, X^{i}) \to \xi(X^{i}) \equiv \oint \frac{d z}{ 2 \pi i z} \xi(z,\bar{z}, X^{i})  $$
The reason is the same as in  \eqref{single-interval-deform}, \ie we only want the mode that moves the cut-off tube instead of deforming it.

 The universal data  regarding the linear response $\delta I_{A,B}$ is encoded in the following ``susceptibility" of shape deformation:
\begin{equation}
	\begin{aligned}
		\frac{\delta I_{A,B}}{\delta \xi(X^{i})} = (i)\oint_{|z|=\tilde{\epsilon}} {d}z \,\  \left\langle \bigg(T_{zz}(z,\bar{z}, X^{i}) + T_{z\bar{z}}(z,\bar{z}, X^{i})\bigg) \Delta \tilde{H}_{A, B}  \right\rangle + h.c. \\
	\end{aligned}
\end{equation}
 which simply picks up the residue of simple poles in $ \langle T_{zz} \Delta \tilde{H}_{A,B} \rangle$, etc. Unlike in the two-dimensional case of the warm-up exercise, strictly speaking the integrands are neither holomorphic nor anti-holomorphic functions of $z $ and $\bar{z}$. However, in the limit $\epsilon \to 0$ the contributions from the integrands can be approximated by corresponding holomorphic or anti-holomorphic functions, see Appendix \ref{sec:solve-the-integral}.

 The basic building blocks for the integrands are three points functions $\langle T \mathcal{O}\mathcal{O} \rangle$. In $ d $ dimensional CFTs, they are completely fixed by conformal symmetry to have the form \cite{Simmons-Duffin2016}:
\begin{equation}
	\left\langle\mathcal{O}\left(X_{1}\right) \mathcal{O}\left(X_{2}\right) T^{\mu \nu}\left(X_{3}\right)\right\rangle=C_{12 T} \frac{H^{\mu \nu}\left(X_{1}, X_{2}, X_{3}\right)}{\left|X_{12}\right|^{2 \Delta-d+2}\left|X_{13}\right|^{d-2}\left|X_{23}\right|^{d-2}},
	\label{TOO}
\end{equation}
\begin{equation}
	\begin{aligned}
		H^{\mu \nu} &=V^{\mu} V^{\nu}-\frac{1}{d} V_{\alpha} V^{\alpha} \delta^{\mu \nu}, \quad V^{\mu} = \frac{X_{13}^{\mu}}{X_{13}^{2}}-\frac{X_{23}^{\mu}}{X_{23}^{2}} \\
		C_{12T}  &=-\frac{d \Delta}{d-1} \frac{1}{S_{d}} C_{12}, \quad S_{d}=\frac{2 \pi^{d / 2}}{\Gamma(d / 2)} .
	\end{aligned}
\end{equation}
where $C_{12}$ is the normalization constant  of the two-point function $ \left\langle \mathcal{O}_{1}\left(x_{1}\right) \mathcal{O}_{2}\left(x_{2}\right) \right\rangle$ which we set to  $ 1 $.  The details of doing integrals are complicated so we package them in Appendix \ref{sec:solve-the-integral} where interested readers can refer to. We will instead make a few general comments and observations below. 
 
  Due to the general form of the modular Hamiltonian $\Delta \tilde{H}_{A,B}$, there are two corresponding integrals for $\delta I_{A,B}$, schematically of the form: 
\begin{equation}
    \begin{aligned}
         \delta {I}_{A,B}^{(1)} \sim &\oint {d z} \; \int {d s} \; k_{1}(s)  \left\langle T \mathcal{O}(X[s]) \mathcal{O}(X[0] ) \right\rangle, \\
       \delta {I}_{A,B}^{(2)} \sim  &\oint {d z} \; \int {d s_{k}} \; \int {d s_{j}} \; k_{2}(s_{k},s_{j})  \left\langle T \mathcal{O}(X[s_{j}+ s_{k}]) \mathcal{O}(X[s_{j}] ) \right\rangle
    \end{aligned}
\end{equation}
 Analogous to $H_{A}$ for the single-interval case  \eqref{single-mH}, $\Delta \tilde{H}_{A,B} $ is also written as modular integral (or double integrals) of local operators. Therefore, according to section \ref{sec:1-interval} we should expect the corresponding simple poles in $z$ to emerge only after completing all the  integrals in modular parameters.
 It turns out the single integral $\int {d s}$ in $\delta I_{A,B}^{(1)}$ does not produce simple poles of $z$, thus  its contribution is zero.  The response comes entirely from the double integral $\int {d s_{k} } \int {d s_{j}}$ in $\delta I_{A,B}^{(2)}$. However, it is slightly surprising that one does not need to complete both integrals for simple poles in $z$ to emerge; only one of the integrals $\int d s_{j}$ is sufficient. The response is then obtained by integrating the  $s_{k}$-dependent residue over the remaining modular parameter. In the end, we obtain the following result:
\begin{equation}
    \begin{aligned}
        \delta I_{A,B} =  N_{\Delta, d} \int {d}^{d-2} X^{i} \,\ \xi(X^{i}) (1 + D)^{-d+1}   \\
    \end{aligned}
    \label{Rindler-result}
\end{equation}
where $ D = \sum_{i=2}^{d-1} (X^{i})^{2} $ and $ N_{\Delta, d} = \Delta \rho^{2 \Delta}  \frac{\sqrt{\pi}}{4^{2\Delta + 1}} \frac{\Gamma(2\Delta+1)}{\Gamma(2\Delta + 3/2)}  \frac{ 2^{d-1} \Gamma\left(\frac{d-1}{2}\right) }{\pi^{d/2-1 / 2}} $. 
 
\subsection{Mapping back to spherical frame}
Now that we have the results in Rindler frame, it is simple to translate them back in the spherical frame and obtain the general results on spheres. Writing \eqref{Rindler-result} in terms of the original deformation field $\zeta$ by \eqref{main-relation}, we have that:
\begin{equation}
	\begin{aligned}
		\delta I_{A, B} =   N_{\Delta, d}  \int {d}^{d-2} X^{i} \,\ {\zeta}(X) C_{I}^{-1}\bigg( 1 +  \frac{C_{I}^{2}}{4R_{A}^{2}} \sum_{i \ge 2} X_{i}^{2} \bigg)  \bigg(1 + \sum_{i \ge 2} X_{i}^{2} \bigg)^{-d+1}   
	\end{aligned}
  \label{I-Rindler-to-sphere}
\end{equation}
The last step is to rewrite \eqref{I-Rindler-to-sphere} in terms of spherical coordinates $(\alpha, e_{i})$ on sphere $A$. The integral measure now takes the form: 
\begin{equation}
	\begin{aligned}
		\int {d}^{d-2}X_{i} 
		&=  \int  {d} {{\Omega}_{d-2}^{Y}} \,\ R_{A}^{d-2} C_{I}^{-(d-2)}  \Big( 1 +  \frac{C_{I}^{2}}{4R_{A}^{2}} \sum_{i \ge 2} X_{i}^{2} \Big)^{d-2} ,
	\end{aligned}
\end{equation}
  From \eqref{cooridinate-relation},  we have that: $$  \sum_{i \ge 2} X_{i}^{2} = \frac{4R_{A}^{2}}{C_{I}^{2}} \frac{1-\cos\alpha}{1+\cos\alpha} $$ 
  Then combining everything, we finally obtain the result for the linear response of mutual information between spheres, for the OPE contribution from scalar primary operator of scaling dimension $\Delta$, it takes the form : 
\begin{equation} \label{eq:main_result}
    \begin{aligned}
       & \delta I_{A, B}^{\Delta, \ell =0} =  \tilde{N}_{d} \; \bigg( \frac{ \partial I_{A,B}^{\Delta, \ell =0} }{\partial \ln \rho } \bigg)  \int {d}{{\Omega}_{d-2}} \,\  \left(\frac{ L^{2}- R_{A}^{2}}{L^{2} + R_{A}^{2} -2 L R_{A} \cos \alpha }\right)^{d-1} {\zeta}(\Omega)
    \end{aligned}
\end{equation}
 where $ \tilde{N}_{d} = \left(R_{A} S_{d-1} \right)^{-1} $ and  we have made the identification:
\begin{equation}
     2\Delta  \rho^{2\Delta} \frac{\sqrt{\pi}}{4^{2\Delta + 1}} \frac{\Gamma(2\Delta+1)}{\Gamma(2\Delta + 3/2)} = \left(\frac{ \partial I_{A,B}^{\Delta, \ell =0} }{\partial \ln \rho }  \right)
\end{equation}
One can plug in (\ref{eq:main_result}) special deformations of the form:   
\be\label{eq:special_def}
 {\zeta}^d (\Omega) =\zeta_0,\;\;  {\zeta}^t(\Omega) =\zeta_0 \cos \alpha
\ee
which generates dilation $ R_{A} \to R_{A} + \zeta_0 $ and translation $L \to L - \zeta_0$ of the sphere $A$. We can compute for these deformations:
\begin{equation} 
    \begin{aligned}
        &\delta_{\zeta^d} I_{A, B}^{\Delta, \ell =0}  = \left( \frac{\zeta_{0}}{ R_{A}} \frac{L^{2} + R_{A}^{2}}{L^{2}-R_{A}^{2}} \right)\left(\frac{ \partial I_{A,B}^{\Delta, \ell =0} }{\partial \ln \rho }\right) \\
       & \delta_{\zeta^t} I_{A, B}^{\Delta, \ell =0}  =  \left(  \frac{ 2 \zeta_{0} L }{L^{2}-R_{A}^{2}}  \right)\left( \frac{ \partial I_{A,B}^{\Delta, \ell =0} }{\partial \ln \rho }\right)
    \end{aligned}
\end{equation}
which agrees with the result obtained by varying $R_{A}$ and $L$ directly in \eqref{MI-Rindler} . 

Eq (\ref{eq:main_result}) is the main result of our paper. The most important feature is a form factor that depends only on the space-time dimension $d$, and is otherwise quite universal, e.g. independent of the scaling dimension $\Delta$. In addition, the ``susceptibility" of shape deformation (refered to as simply susceptibility from now on) is only proportional to powers of the distance between the center of small ball $ B $ and deformation point $\Omega$ on $ A $: 
\be
\frac{\delta I^{\Delta,\ell=0}_{A, B} }{\sqrt{h(\Omega)}\delta \zeta(\Omega)} \propto |Y_B -Y_\Omega|^{-2(d-1)}
\ee
Therefore, the shape deformation only induces a response that corresponds to the ``zero-mode" in a sense that we will discuss later in section \ref{sec:extremality}. 

At this point, these observations only pertains to a particular type of contribution, i.e. those from scalar primary operators. In principle, the susceptibility could still depend on the spin $\ell$ of the primary operator forming the bi-local OPE channel. It is therefore worth studying the effect of including spins to the computation of shape response. Conducting an exhaustive investigation of all spinning contributions is however beyond the scope of our goal, instead we study the case with spin-1 primary operator, and observe if the corresponding susceptibility is modified. 

\subsection{Including spins: case study for spin-1 contributions}\label{sec:spin-1}
We now study the OPE contribution to mutual information $I_{A,B}$ from the bi-local operator consisting of two spin-1 primary operators with conformal dimension $\Delta$: 
\be
\left\langle \Sigma^{B}_n\right \rangle^{-1}\times \Sigma^{B}_n \rightarrow \frac{1}{2}( 2R_B)^{2\Delta}\sum_{j\neq k} c^{\mu\nu}_{j-k}\mathcal{O}_\mu^{(j)}\mathcal{O}_\nu^{(k)}
\ee
The goal is to compute the linear response of this contribution under shape deformation. Similar to the scalar case, the OPE coefficients $c^{\mu\nu}_{j-k}$ for spinning primary operator is related to the thermal correlator of spinning operators in the hyperbolic space $\mathds{H}_{d-1}$ at inverse temperature $T^{-1} = 2\pi n$ using (\ref{eq:hyperbolic}): 
\bea
 c^{\mu\nu}_{j-k} &=& \lim_{r\to \infty}\Omega_i^{\Delta-1}\Omega_j^{\Delta-1}\left(\frac{dZ_j}{dr}\right)^\rho_\alpha \left(\frac{dZ_k}{dr}\right)^\sigma_\beta G^{-1}(r)^\mu_\rho G^{-1}(r)^\nu_\sigma \left\langle \mathcal{O}^\alpha(Z_j)\mathcal{O}^\beta(Z_k)\right\rangle_{\mathds{S}_{(2\pi n)}\times\mathds{H}_{d-1}} \nonumber\\
&=& G^{\mu\nu}_n(\tau_j-\tau_k),\;\;\tau_i = (2i+1)\pi
\eea
A systematic treatment of replica trick for OPE into spinning primaries as well as their contributions to the mutual information can be found in \cite{Casini2021Sep}, see also \cite{ChenBin2017,Long2016}. In order to compute its shape response using the entanglement first law, we need to know its OPE contribution to the modular Hamiltonian. Although the derivation in \cite{Faulkner2021Aug} was applied explicitly to the scalar case \eqref{mH-two-sphere}, the properties used in the derivation (KMS condition, translation invariance in replica indices) are valid for $c^{\mu\nu}_{j-k}$ as well. As a result, one can simply substitute $c_1(is)\to c^{\mu\nu}_1(is)$ and thus obtain the spin-1 contribution to the modular Hamiltonian: 
\bea\label{eq:spin_1_mH}
\Delta H^{\Delta,\ell=1}_{A,B} &=& -{\rho^{2 \Delta}}  \int_{-\infty}^{\infty} {d s} \,\  k^{\mu\nu}_{1}(s) \rho_{A}^{i s} \mathcal{O}_\mu(-X_B) \rho_{A}^{-i s} \mathcal{O}_\nu(X_B)\nonumber\\
&+& \frac{i \rho^{2 \Delta}
}{2 \pi }
\int_{-\infty}^{\infty} {d s_{j} d s_{k}} \,\ k^{\mu\nu}_{2}(s_{j}, s_{k}) \rho_A^{-i(s_j+s_k)}\mathcal{O}_\mu(-X_B)\; \rho_A^{is_k}\mathcal{O}_\nu(-X_B)\;\rho_A^{is_j} + \delta H_{II}\nonumber\\
k^{\mu\nu}_{1}(s) &=& \frac{1}{4\cosh^2(s/2)}c^{\mu\nu}_1(-is+\pi) \nonumber\\
k^{\mu\nu}_{2}(s_{j}, s_{k}) &=& \frac{1}{4 \cosh^{2} (s_{j}/2)} \left( \frac{1}{e^{s_{k}+ i \epsilon}-1} + \frac{1}{e^{s_{k}+s_{j}}+1} \right) c^{\mu\nu}_{1}(-is_{k} + \epsilon) 
\eea
The coefficient $c_1^{\mu\nu}(is)$ is the analytically continued correlator in hyperbolic space at $n=1$, which is conformally related to flat-space correlator: 
\bea
&&c^{\mu\nu}_1\left(is-is'\right) = \left\langle \mathcal{O}^\mu(is,0)\mathcal{O}^\nu(is',0)\right\rangle_{\mathds{S}_{2\pi} \times \mathds{H}_{d-1}} \nonumber\\
&=& \lim_{r\to\infty}\left(\frac{\partial r(is)}{\partial Z}\right)^\mu_\alpha \left(\frac{\partial r(is')}{\partial Z}\right)^\nu_\beta \left\langle \mathcal{O}^\alpha(r(is))\mathcal{O}^\beta(r(is'))\right\rangle_{\mathds{R}^d} 
\eea
The trajectories $r(is)$ are the images of hyperbolic time translation $\tau\to\tau+is$ in $\mathds{R}^d$ starting at $r(0)=r$. They are nothing but the modular-flow trajectories: 
\be
r(s) =\frac{R\left((r^2-R^2)\sinh{(s)},-2R r_i\right)}{2\left(r^2\sinh^2{\left(\frac{s}{2}\right)}-R^2 \cosh^2{\left(\frac{s}{2}\right)}\right)} 
\ee
We need to take the $r\to \infty$ limit to extract the OPE coefficient, in which case the trajectories simplify to:
\be
\lim_{r\to\infty}r(s)=  \left(R\coth{\left(\frac{s}{2}\right)},\vec{0}\right)
\ee
We can then obtain that: 
\be\label{eq:spin_1_OPE}
c^{\mu\nu}_1(is) = c_1(is) \left(\eta^{\mu\nu}-2\hat{n}^\mu\hat{n}^\nu\right)  
\ee
where $c_1$ is the scalar coefficient and $\hat{n}^\mu=n^\mu/(n\cdot n)^{1/2}$ is the normalized time-like vector pointing from the center of the sphere to the tip of its null cone. In our case $\eta^\mu = \delta^\mu_0$ but the formula works more generally for boosted spheres. The dependence on $\hat{n}^\mu$ in the OPE coefficients encodes the orientation of the twist operator $\Sigma^I_n$, as in the usual OPE expansion. Plugging (\ref{eq:spin_1_OPE}) into (\ref{eq:spin_1_mH}), the contribution of spin-1 primary to mutual information can be found to be $d$ times that of scalar primary: 
\be\label{eq:spin_1_MI}
I^{\Delta,\ell=1}_{A,B} = d \rho^{2 \Delta}  \frac{\sqrt{\pi}}{4^{2\Delta + 1}} \frac{\Gamma(2\Delta+1)}{\Gamma(2\Delta + 3/2)} 
\ee
This agrees with results in \cite{Casini2021Sep, ChenBin2017, Long2016}.  

Now one can proceed and compute the linear response of (\ref{eq:spin_1_mH}) to shape deformation using the entanglement first law. The strategy is identical to the scalar computation, and one still finds that only the double integral term in (\ref{eq:spin_1_mH}) produces a response. We shall not repeat the details and instead zoom into the key computation, which is the following integral: 
\be
\delta I^{(2)}_{A, B} \propto \oint dz \int^{\infty}_{-\infty} ds_j ds_k k^{\mu\nu}_2(s_j,s_k)\left\langle T_{zz}(z,\bar{z};X^i) \rho_A^{-i(s_j+s_k)}\mathcal{O}_\mu(-X_B)\; \rho_A^{is_k}\mathcal{O}_\nu(-X_B)\;\rho_A^{is_j} \right\rangle \nonumber
\ee
For spinning operators, the modular flow is also accompanied by a tensor factor:
\bea 
\rho_A^{-is}\mathcal{O}_\mu(X)\rho_A^{is} = \Omega_s^{\Delta-1} R(s)^\alpha_\mu\; \mathcal{O}_\alpha\left(X[s]\right)
\eea
In the Rindler frame where we evaluate the integral, it takes the form of the boost:
\be
R(s)^\alpha_\mu = e^{s}\delta^+_\mu \delta^\alpha_+ + e^{-s}\delta^-_\mu \delta^\alpha_-+ \delta^{\alpha}_{\mu}|_{\alpha,\mu\geq 2}
\ee
For spinning operators, one also needs the tensor structure $Q_{zz\mu\nu}$ appearing in the relevant three-point function: 
\be
\left\langle \mathcal{O}_{\mu}(x_1)\mathcal{O}_{\nu}(x_2)  T_{zz}(x_3) \right\rangle = \frac{ Q_{\mu\nu zz}(x_{12},x_{13},x_{23})}{|x_{12}|^{2\Delta-d}|x_{13}|^{d-2}|x_{23}|^{d-2}} 
\ee
Conformal symmetry restricts the form of the tensor structures $Q$ to be a linear combination of 4 basic building blocks, which can be most compactly encoded using the embedding space coordinates $P$ and the corresponding auxiliary null vector $T$ \cite{Costa2011}: 
\bea 
Q &=& a_1 V_{1,23}V_{2,31}V_{3,12}^2+a_2 V_{3,12}^2H_{12}\nonumber\\
&+& a_3 \left(V_{1,23}H_{23}+V_{2,31}H_{13}\right)V_{3,12} +a_4 H_{13}H_{23}
\eea 
The building blocks $V_{i,jk}$ and $H_{ij}$ are defined by: 
\bea
V_{i,jk} &=& \frac{(T_i\cdot P_j)(P_i\cdot P_k)-(T_i\cdot P_k)(P_i\cdot P_j)}{(P_j\cdot P_k)}\nonumber\\
H_{ij} &=& -2\left[(T_i\cdot T_j)(P_i\cdot P_j)-(T_i\cdot P_j)(T_j\cdot P_i)\right] 
\eea
To read out the components $Q_{\mu\nu zz}$, one first re-write the contractions in terms of physical coordinates $x$ and the corresponding auxiliary vector $t$ using: 
\bea
T_i\cdot T_j \to t_i\cdot t_j,\;\;P_i\cdot P_j \to -\frac{1}{2} x_{ij}^2,\;\;P_i\cdot T_j \to t_j\cdot x_{ij} 
\eea
and then extract the coefficient of $t^\mu_1 t^\nu_2 t^z_3 t^z_3$. The computation proceeds more tediously but is otherwise identical to the scalar case, so we spare the details and simply write the result in Rindler frame: 
\bea\label{eq:spin_1_response}
  \delta I^{\Delta,\ell = 1}_{A, B}  &=&  N_{\Delta, d}^{\ell =1} \int {d^{d-2} X} \;  \left(1+\sum_{i\geq 2} X_i^2\right)^{-(d-1)} \zeta(X)\nonumber\\ 
   N_{\Delta, d}^{\ell =1} &=&    \frac{ 2^{d-1}  }{S_{d-1}}\left( \frac{ \partial I_{A,B}^{\Delta,\ell =1 }}{\partial  \ln \rho }\right)  \frac{(d-1)S_{d}}{ d^{2}\Delta} \left(a_1-a_2 d-2 a_3+2a_4\right)
\eea
In the spherical frame, we would then have that: 
\begin{equation} \label{eq:main_result-spin-1}
    \delta I^{\ell = 1}_{A, B} =  \left(\frac{2^{1-d}}{R_A}\right)N_{\Delta, d}^{\ell = 1} \int {d \Omega_{d-2}} \;  \left(\frac{ L^{2}- R_{A}^{2}}{L^{2} + R_{A}^{2} -2 L R_{A} \cos \alpha }\right)^{d-1}  {\zeta}(\Omega)
\end{equation}
We can see that the susceptibility contains the same form factor as the scalar contribution in (\ref{eq:main_result}). The coefficients $a_i, i=1,...,4$ are generally theory-dependent. In our case, they are further constrained by the conservation of $T_{\mu\nu}$ and its Ward identity. This will reduce the number of independent coefficients, but there will still be free parameters that are theory-dependent. In particular, consistency with (\ref{eq:spin_1_MI}) for special deformations (\ref{eq:special_def}) then requires that: 
\be \label{eq:spin_1_consistency}
a_1-a_2 d-2 a_3+2a_4 = \frac{d^2\Delta}{(d-1)S_d} 
\ee
so that: 
\be\label{eq:spin_1_main}
\delta I^{\Delta,\ell=1}_{A,B}  =   \tilde{N}_{d} \; \left( \frac{ \partial I_{A,B}^{\Delta,\ell =1 }}{\partial  \ln \rho } \right) \int {d}{{\Omega}_{d-2}} \,\  \left(\frac{ L^{2}- R_{A}^{2}}{L^{2} + R_{A}^{2} -2 L R_{A} \cos \alpha }\right)^{d-1} {\zeta}(\Omega)
\ee
For the special case of $\mathcal{O}_\mu$ being a conserved current with conformal dimension $\Delta = d-1$, the coefficients can be more explicitly written in \cite{Osborn1994, Hofman2016, Dymarsky2019} as: 
\bea
a_1 &=& (2-3d-4d^3\gamma) C,\;\; a_2 = (1-2d-4d^2\gamma)C\nonumber\\
a_3 &=& -2d(1+4\gamma)C,\;\;a_4=2\left(\frac{1}{d-2}-4\gamma d\right)C,\;\;C = \frac{d(d-2)}{2(d-1)^2S_d}
\eea
where we have normalized the current two-point function. In this case one can check that the free parameter $\gamma$ cancels out and (\ref{eq:spin_1_consistency}) is indeed satisfied. In principle, (\ref{eq:spin_1_consistency}) should be derivable in CFTs from the conformal ward identity directly. In our context it is imposed for the shape dependence of mutual information, which is obtained from $\langle T_{\mu\nu}\mathcal{O}_\alpha\mathcal{O}_\beta\rangle$, to be consistent with the change in mutual information between spheres $\delta I \sim \delta \langle \mathcal{O}_\alpha\mathcal{O}_\beta\rangle$ under conformal transformations (\ref{eq:special_def}) -- this is essentially the conformal ward identity. Since the scalar and spin-1 contributions (\ref{MI-Rindler}, \ref{eq:spin_1_MI}) to the mutual information between spheres are themselves highly universal -- they only depend on the conformal ratio, this also explains partially the universality we observe in their shape-dependences.

We conclude that for the bi-local OPE contribution from spin $\ell =1$ primary operators, the susceptibility under shape deformation reveals the same form factor as the scalar result (\ref{eq:main_result}). Therefore, it is worth conjecturing that the susceptibility is of the universal form: 
\be 
\delta I^{\Delta,\ell}_{A, B} = \tilde{N}_{d} \;  \left(\frac{ \partial I^{\Delta,\ell}_{A, B} }{\partial \ln \rho }  \int {d}{{\Omega}_{d-2}}\right) \,\  \left(\frac{ L^{2}- R_{A}^{2}}{L^{2} + R_{A}^{2} -2 L R_{A} \cos \alpha }\right)^{d-1} {\zeta}(\Omega) 
\ee
for all bi-local OPE contributions from primary operators, i.e. independent of both their conformal dimension $\Delta$ and spin $\ell$. 

\section{Extremization of mutual information in the OPE limit}\label{sec:extremality}
In this section, we try to unpack some implications of (\ref{eq:main_result}). In the presence of symmetries, the response theory can be diagonalized into eigenmodes of the symmetry generators. For example, in \cite{Mezei2015Feb,Mezei2015Feb02} the linear response under shape deformation of entanglement entropy of spherical sub-regions in CFT vacuum has been found to only invoke the zero mode of spherical harmonics (i.e. the average) multiplied by a cut-off dependence -- an  indication for the extremization of its universal part on spheres: 
\be
\delta S \propto \delta^{-(d-2)} \int_{S_{d-2}} d\Omega_{d-2} \zeta(\Omega) 
\ee
In this case the spherical harmonics should emerge as the natural candidate for mode decomposition of the response, this is due to the rotational symmetry of the set up. To clarify more, the role of rotational symmetry has two aspects: firstly, it is the symmetry of the underlying CFT, and in particular of its vacuum; secondly, the entangling surface (thus the defect twist operator $\Sigma_n$ inserted when computing via the replica trick) is kept invariant under the action of symmetry transformation. To be more precise, the vacuum is symmetric under rotations about any point; the spherical defect is kept invariant by rotations about a particular point as the center -- the ``little group" of $\Sigma_n$.   

\begin{figure}[h]
     \centering
     \includegraphics[scale=0.55]{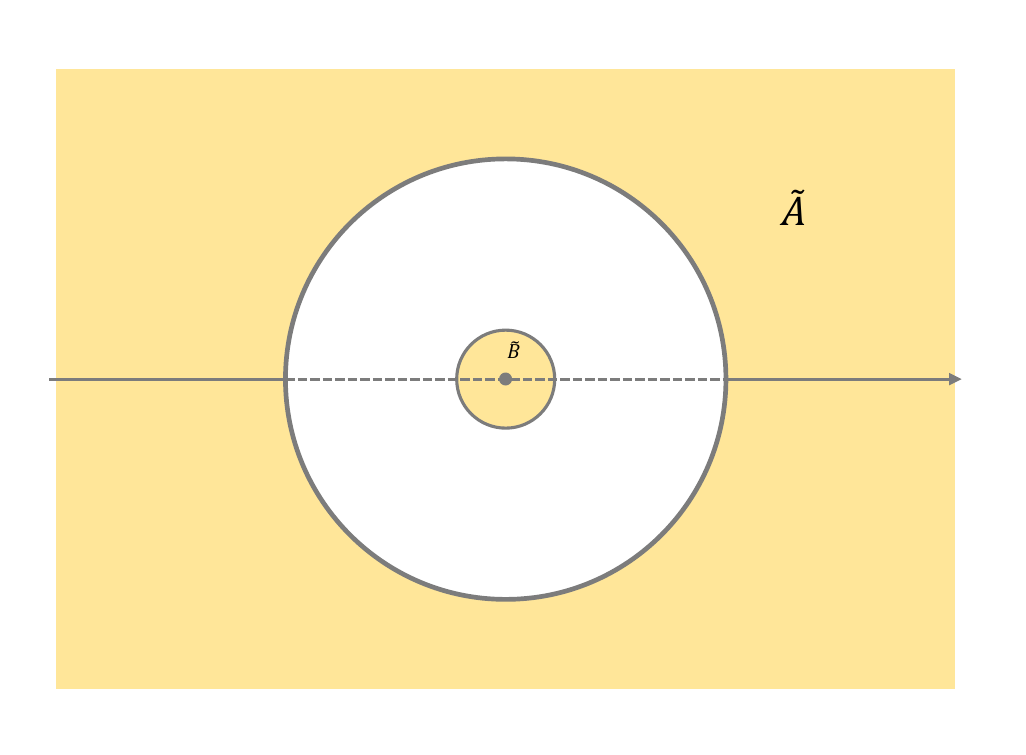}
     \caption{Conformally equivalent configuration of two concentric spheres. } 
     \label{fig:concentric-ball}
 \end{figure}

Now with regard to mutual information between two spherical sub-regions $A$ and $B$, the ``little group" is generically empty in the presence of two spherical defects $\Sigma^A_n$ and $\Sigma^B_n$ for vacuum in generic relativistic QFTs. However for CFTs we still have a non-trivial little group, i.e. symmetries of the CFT vacuum that keep both sub-regions invariant. The crucial ingredient is a special conformal transformation $\mathcal{K}$ that maps two spheres to be positioned concentrically (see Fig \ref{fig:concentric-ball}).

More concretely, we denote the centers of two spheres by $Y_A$ and $Y_B$ respectively, separated by distance $L$. The special conformal transformation $\mathcal{K}$ consists of two inversions across carefully chosen origins. The first inversion $\mathcal{I}_1$ across the origin $Y_0$ needs to make $B$ cover $A$ co-centrically. We can parametrize $Y_0$ by $\lambda$: 
\bea
Y_0= \lambda Y_A + (1-\lambda) Y_B,\;\; 0<\lambda<1
\eea
The inversion $\mathcal{I}_1$ maps the two spheres to be centered at: 
\bea
\tilde{Y}_A &= & Y_0 - \frac{(1-\lambda)\left(Y_B-Y_A\right)}{(1-\lambda)^2 L^2-R_A^2}\nonumber\\
\tilde{Y}_B &= & Y_0 - \frac{\lambda\left(Y_B-Y_A\right)}{R_B^2-\lambda^2 L^2}
\eea
Solving for the co-centricity condition $\tilde{Y}_A =\tilde{Y}_B$ gives that:
\be
\lambda = \frac{1}{2}\left(1+r_B^2-r_A^2-\sqrt{\left(1+r_B^2-r_A^2\right)^2-4r_B^2}\right),\;\;r_{A,B} = \frac{R_{A,B}}{L} 
\ee
so that the common center $\tilde{Y}_c = \tilde{Y}_{A,B}$ is located at:
\bea
\tilde{Y}_c &=& \tilde{\lambda}Y_A + (1-\tilde{\lambda})Y_B\nonumber\\
\tilde{\lambda}&=& \lambda +\frac{1}{L^2}\left(\left(1+r_B^2-r_A^2\right)^2-4r_B^2\right)^{-1/2}  
\eea
The second step $\mathcal{I}_2$ is then another inversion across $\tilde{Y}_c$. Combining both steps, we obtain the required special conformal transformation $\mathcal{K}=\mathcal{I}_2\circ \mathcal{I}_1$:
\be\label{eq:SCT}
\mathcal{K}:\;\;Y^\mu \to \frac{(Y^\mu-a^\mu) - b^\mu (Y-a)^2}{1-2(Y-a)\cdot b + b^2 (Y-a)^2},\;\; a^\mu = Y^\mu_c,\;\;b^\mu = \tilde{Y}^\mu_c -Y^\mu_c
\ee
and the radii of the co-centric images $\tilde{A}$ and $\tilde{B}$ of the two spheres become: 
\be
R_{\tilde{A}} = \frac{(1-\lambda)^2 L^2-R_A^2}{R_A},\;\;R_{\tilde{B}} = \frac{R_B^2-\lambda^2 L^2}{R_B} 
\ee
The symmetries of the conformal vacuum that preserves both spheres $A$ and $B$ can then be constructed using $\mathcal{K}$: 
\be\label{eq:symmetry_mapping}
\hat{\mathcal{R}}_{A,B}(\Omega) = \mathcal{K}^{-1}\circ \hat{\mathcal{R}}(\Omega) \circ \mathcal{K},\;\;\tilde{R}(\Omega) \in \text{SO}(d-2) 
\ee
The natural modes of expansion $\Phi^{(\ell,m)}_{A,B}$ on $A$ associated with the symmetry action $\hat{\mathcal{R}}_{A,B}$ can then be related to the usual spherical harmonics $\Phi^{(\ell,m)}$ associated with ordinary rotations $\hat{\mathcal{R}}$ by pulling back the special conformal transformation: 
\be\label{eq:pull_back}
\Phi^{(\ell,m)}_{A,B} = \Phi^{(\ell,m)}\circ \mathcal{K}
\ee 
where $\mathcal{K} : (\alpha, e_i) \to (\tilde{\alpha},\tilde{e}_i)$ now maps between angular variables on $A$ and $\tilde{A}$, whose form we can deduce by restricting (\ref{eq:SCT}) on the two spheres: 
\bea\label{eq:angular_map_full}
\tan{\tilde{\alpha}} &=& \frac{R_A\left(K -R_A\cos{\alpha}\right)-\tilde{K}\sqrt{R_A^2+K^2 -2R_A K\cos{\alpha}-\tilde{K}^2\sin^2{\alpha}}}{\left(\tilde{K}\sin{\alpha}+R_A \cos{\alpha}-K\right)\left(\tilde{K}-R_A \cot{\alpha}+K\csc{\alpha}\right)}\nonumber\\
\tilde{K}&=&(\lambda-\tilde{\lambda})\left[(1-\lambda)^2 L^2-R_A^2\right]L\nonumber\\
K&=& (1-\lambda)L,\;\;\tilde{e}_i = e_i
\eea 
Furthermore, the definition of inner products between these modes $\Phi^{(\ell,m)}_{A,B}$ come with an additional change of integration measures due to the conformal factor $\mathcal{J}(\Omega)\sim |\partial \mathcal{K}|$: 
\be
\left\langle \Phi^{(\ell,m)}_{A,B},\; \Phi^{(\ell',m')}_{A,B}\right\rangle = \int_{S^A_{d-2}} d\Omega_{d-2} \mathcal{J}(\Omega)^{d-2} \left[\Phi^{(\ell,m)}_{A,B}(\Omega)\right]^* \times \Phi^{(\ell',m')}_{A,B} (\Omega) 
\ee 
In our case, the deformation field is actually the contravariant radial component of a normal pointing vector field $\zeta(\Omega) =\zeta^r(\Omega)$, so its decomposition into harmonics modes using pull back (\ref{eq:pull_back}) of the special conformal transformation $\mathcal{K}$ takes the form:
\bea \label{eq:deform_decom}
\zeta^r(\Omega) &=& \sum_{\ell,m}\zeta^{(\ell,m)} \left[\partial \mathcal{K}(\Omega)^{-1}\right]^r_{\tilde{r}} \left(\Phi^{(\ell,m)}\circ \mathcal{K}\right)(\Omega)\nonumber\\
&=& \mathcal{J}^{-1}(\alpha,e_i) \sum_{\ell,m} \zeta^{(\ell,m)} \Phi^{(\ell,m)}_{A,B}(\Omega)
\eea
The full mapping (\ref{eq:angular_map_full}) between angular variables simplifies in the OPE limit $R_B\to 0$ that we have been working with so far:
\bea
\tan{\tilde{\alpha}} \to \frac{\left(R_A^2-L^2\right)\sin{\alpha}}{\left(L^2+R_A^2\right)\cos{\alpha}-2L R_A},\;\;\tilde{e}_i = e_i 
\eea
We can then obtain the conformal factor $\mathcal{J}$ as: 
\be
\mathcal{J}(\alpha,e_i)=\frac{1-r_A^2}{1+r_A^2-2r_A\cos{\alpha}}
\ee
which is what appears in our main result (\ref{eq:main_result}). At this stage, we can re-write (\ref{eq:main_result}) in terms of the above mode decomposition: 
\bea
&&\delta I^{\Delta,\ell=0}_{A,B} \propto  \int d\Omega_{d-2} \left(\frac{1-r_A^2}{1+r_A^2-2r_A \cos{\alpha}}\right)^{d-1}\zeta(\Omega)\nonumber\\ 
&=&\sum_{\ell,m} \zeta^{(\ell,m)} \int  d\Omega_{d-2} \mathcal{J}(\Omega)^{d-2}\; \Phi^{(\ell,m)}_{A,B}(\Omega) \propto \sum_{\ell,m}\zeta^{(\ell,m)}\left\langle \Phi^{(0,0)},\Phi^{(\ell,m)}\right\rangle = \zeta^{(0,0)}
\eea
We see that under an arbitrary shape deformation $\zeta(\Omega)$, the linear response of mutual information in the OPE limit only picks up the zero-mode component $\zeta^{(0,0)}$ of $\zeta(\Omega)$ according to (\ref{eq:deform_decom}). In other words, the OPE limit of $I_{A,B}$ extremizes against all non-zero modes of shape deformation: 
\be
\frac{\delta I^{\Delta,\ell=0}_{A,B}}{\delta \zeta^{(\ell,m)}}=0,\;\;\;(\ell,m)\neq (0,0) 
\ee  
We can understand better the zero-mode nature of the response $\delta I^{\Delta,\ell=0}_{A,B}\propto \zeta^{(0,0)}$ from a symmetry point of view. When acting on the sphere $A$, the zero-mode component $\zeta^{(0,0)}$ of deformation: 
\bea
\zeta^{0,0} (\Omega) &=& \zeta^{(0,0)} \times \mathcal{J}^{-1}\left(\alpha,e_i\right)\nonumber\\
&\underset{R_B\to 0}{\approx}&\; \zeta^{(0,0)} \left(\frac{1+r_A^2}{1-r_A^2}\right)- \zeta^{(0,0)}\left(\frac{2r_A\cos{\alpha}}{1-r_A^2}\right)
\eea
implements simultaneously a dilation of $R_A$ and a translation of the center $Y_A$ by: 
\be \label{eq:zero_mode_action}
R_A \to R_A + \zeta^{(0,0)}\left(\frac{1+r_A^2}{1-r_A^2}\right),\;\; Y_A = Y_A - \zeta^{(0,0)}\left(\frac{2 r_A}{1-r_A^2}\right)\hat{\mathcal{L}},\;\;\hat{\mathcal{L}}= \frac{Y_B-Y_A}{|Y_B-Y_A|}
\ee
Using (\ref{eq:SCT}, \ref{eq:symmetry_mapping}) one can write down a one-parameter $\chi$ family of invariant spheres under the group actions of $\hat{\mathcal{R}}_{A,B}$, whose radius and center $(\chi,Y_\chi)$ we parametrize by: 
\be
R_\chi = \chi,\;\; Y_\chi = Y_A - L_\chi \hat{\mathcal{L}} 
\ee
They provide a $\hat{\mathcal{R}}_{A,B}$-symmetric foliation of space. In the OPE limit $R_B\to 0$, the spheres are defined by satisfying the relation (see Fig \ref{fig:non-concentric-spheres}): 
\begin{figure}[h]
	\centering
	\includegraphics[scale=0.55]{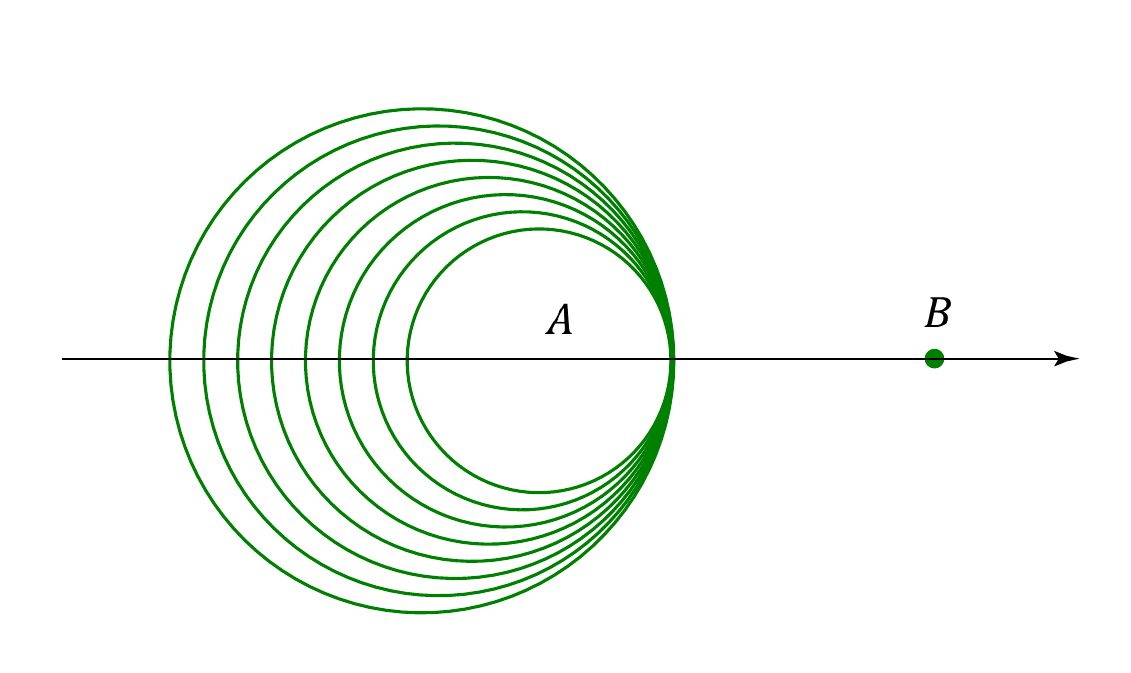}
	\caption{The one-parameter $\chi$ family of invariant spheres in the OPE limit $R_B\to 0$} 
	\label{fig:non-concentric-spheres}
\end{figure}

\be\label{eq:foliation}
\frac{L_\chi}{L_\chi^2 -\chi^2} = \frac{L}{L^2-R_A^2}\; \rightarrow \; L_\chi = \frac{L}{2}\left(1-r_A^2 \pm \sqrt{(1-r_A^2)^2+4 r_\chi^2}\right),\;\;r_\chi = \chi/L
\ee
Zooming near the original sphere $A$ with: 
\be
\chi = R_A+\left(\frac{1+r_A^2}{1-r_A^2}\right)\delta,\;\; \delta \ll R_A
\ee 
and picking the (+) branch of (\ref{eq:foliation}) one can see that the mode of deformation mapping between these invariant spheres: 
\be
R_A \to R_A + \delta \left(\frac{1+r_A^2}{1-r_A^2}\right),\;\; Y_A \to Y_A  - \delta\left(\frac{2r_A}{1-r_A^2}\right) \hat{\mathcal{L}}
\ee 
corresponds to precisely the zero-mode in (\ref{eq:zero_mode_action}). In other words, at leading order under shape deformation, the OPE limit of mutual information is only affected by modes that map between invariant spheres under $\hat{\mathcal{R}}_{A,B}$.

\section{Discussions}\label{sec:discuss}
In this paper, we computed the linear response of mutual information $I_{AB}$ between two spheres $A$ and $B$ under shape deformation on one of the spheres $A\to A+\zeta$. The calculation is based on the OPE of the mutual information and the corresponding modular Hamiltonian in the limit $R_B\to 0$. This enabled us to study the linear response of each OPE contribution of $I_{AB}$ separately. The main result (\ref{eq:main_result}) of our calculation shows that for the possible dominant contribution from a primary scalar operator of conformal dimension $\Delta$, its susceptibility under shape deformation is universal: 
\be
\frac{\delta I^{\Delta,\ell=0}_{A,B}}{\sqrt{h(\Omega)}\delta \zeta(\Omega)} \propto |Y_\Omega-Y_B|^{-2(d-1)}
\ee 
in the sense that it is independent of the conformal dimension $\Delta$. Regarding whether such universality could be spoiled in other possibilities of dominant OPE contributions, e.g. from primary operators with spins, we did a case study for the spin $\ell =1$ contribution and computed its linear response. It is found that the universal form of (\ref{eq:main_result}) persists in this case. It is therefore tempting to speculate that (\ref{eq:main_result}) might be valid for contributions from primary operators with arbitrary $\Delta$ and $\ell$, and thus describes the universal linear response of $I_{AB}$ in the OPE limit. We provided a symmetry perspective to understand the implication of (\ref{eq:main_result}). For the CFT vacuum, there exists a group $\hat{\mathcal{R}}_{A,B}$ of symmetries isomorphic to the rotation group which leaves both the state and the sub-regions $A$ and $B$ invariant. The deformation vector fields $\zeta(\Omega)$ therefore admit a natural mode decomposition associated with the symmetry group $R_{A,B}$: 
\be
\zeta(\Omega) = \sum_{\ell,m} \zeta^{(\ell,m)} \mathcal{J}(\Omega)^{-1} \Phi^{(\ell,m)}_{A,B}(\Omega)
\ee
under which our result (\ref{eq:main_result}) takes the form: $\delta_\zeta I_{AB} \propto \zeta^{(0,0)}$. This implies that for $I_{AB}$ between spherical sub-regions $A$ and $B$, its OPE limit extremizes over all deformations of $A$ that breaks the $\hat{\mathcal{R}}_{A,B}$ symmetry of the configuration.    We end the paper by discussing a few points and propose some related future directions. 
\\
\begin{itemize}[leftmargin=*]
    \item \textit{\textbf{Order of integrations: additive structure in $\delta I_{AB}$}}
    
    Our calculation relied on the application of the entanglement first law and shape perturbation theory. Schematically, the linear response is extracted from contour integrals of the form:
    \be\label{eq:1st_law_simp}
    \delta S/\delta \zeta(x) \propto \oint_{z\sim z_0} dz \left\langle T_{zz}(z,\bar{z},x) H \right\rangle,\;\; (z_0, x) \in \partial A
    \ee
    i.e. finding the residue of the simple pole $(z-z_0)^{-1}$ in the correlator between stress tensor and modular Hamiltonian. How the simple poles are actually produced could reveal a deeper aspect of the modular Hamiltonian. In the cases that one can deal with, the modular Hamiltonians are generally expressed as a finite dimensional integral of few-body local operators: 
    \be \label{eq:mH_gen}
    H = \int ds_1,...,ds_n\; H\left(s_1,...,s_n\right) \equiv \int d\vec{s}\; H\left(\vec{s}\right)
    \ee 
    Such a form could be exact and is a result of symmetry, as in the case of spherical subregion in CFT vacuum; or in our case it could be the leading order contribution of the OPE expansion. In either scenario, the order of integrations is important for getting the correct answer. In particular, it means that schematically: 
    \be\label{eq:int_order}
    \delta S/\delta \zeta \propto \oint dz \int d\vec{s} \left\langle T_{zz}(z) H\left(\vec{s}\right) \right\rangle  \cancel{\propto} \int d\vec{s} \oint dz\left\langle T_{zz}(z) H\left(\vec{s}\right) \right\rangle
    \ee
    In other words, the integral (\ref{eq:mH_gen}) does not respond to shape deformations additively, i.e. the integrand does not directly produce simple poles in $z$ in (\ref{eq:1st_law_simp}), e.g. by OPE with $T_{zz}$, whose residues are then integrated. One should take this as a manifestation about the non-local nature of the entanglement first law, even when the modular Hamiltonian appears as integrals of few-body local operators. The entanglement structure is organized non-locally, whose response to shape deformation is capture by (\ref{eq:mH_gen}) only collectivelly. Therefore, (\ref{eq:int_order}) should be the natural order of integration, and one can expect simple $z$ poles in (\ref{eq:1st_law_simp}) to emerge only after all the integrals of $d\vec{s}$. 
    
    In this aspect, the OPE of $\Delta H_{AB}$ responds to shape deformation in a slightly special way. The linear response is produced by the double integral term:
    \be 
    \Delta H_{AB} \supset  \int ds_j ds_k\;\Delta H^{(2)}_{AB}(s_j,s_k)
    \ee 
    in (\ref{mH-two-sphere}), where we recall: 
    \be
    \Delta H^{(2)}_{AB}(s_j,s_k) \propto \mathcal{O}_{L}(-is_{k} - i s_{j}) \mathcal{O}_{L}(-is_{j})
    \ee
    In this case, one of the integrals (in $s_j$) is sufficient to produce simple poles in $z$. This then allows us to partially switch the order of integration and write:
    \bea
    &&\delta I_{AB} \propto  \oint dz \left(\int ds_j ds_k \left\langle T_{zz}(z) \Delta H_{AB}(s_j,s_k)\right\rangle\right) \nonumber\\
    &=& \int d s_k \left(\oint dz \int ds_j \left\langle T_{zz}(z) \Delta H_{AB}(s_j,s_k)\right\rangle\right) = \int d s_k\; ``\delta I_{AB}(s_k)"
    \eea 
    We interpret this as follows. In terms of response to shape deformation, the modular Hamiltonian (\ref{mH-two-sphere}) contributes additively in $s_k$ which measures the relative modular flow between the bi-local operator $\mathcal{O}(-is_j-is_k)\mathcal{O}(-is_j)$ in the integrand. One is then naturally led to wonder whether the additive behavior of $s_k$ bears more insights into the entanglement structures relevant for mutual information in the OPE limit. In the future, it would be interesting to examine this deeper, for example by probing the additive behavior from other perspectives.  
    
    \item \textit{\textbf{Universality of (\ref{eq:main_result}) among all OPE contributions?}}
    
    The main result (\ref{eq:main_result}) of the paper exhibits a high degree of universality. Its form factor only depends on the space-time dimension $d$. Furthermore, one obtains the same form factor when the primary operator $\mathcal{O}$ forming the bi-local OPE contribution is a scalar or have spin $\ell =1$. We had not planned to perform an extensive check of the validity of (\ref{eq:main_result}) among all OPE contributions. This in particular could include: 
    \begin{enumerate}
        \item \label{item:check1} bi-local operators formed by higher spin $\ell \geq 2$ primary operators;
        \item \label{item:check2} bi-local contributions from descendent operators; 
        \item \label{item:check3} multi-local channel OPE contributions of the form:
        \be
        \Sigma^n_B \rightarrow c_{i_1,...,i_n} \mathcal{O}^{(i_1)}...\mathcal{O}^{(i_n)} \nonumber
        \ee
    \end{enumerate}
    In retrospect, the form factor in (\ref{eq:main_result}) could have been fixed by symmetry if we accept the following two points regarding the linear responses: (i) it transforms covariantly under conformal transformations; (ii) in the conformal frame where $\tilde{A}$ and $\tilde{B}$ are co-centric (see section {\ref{sec:extremality}}), it is rotationally invariant on $\tilde{A}$. The second point is the dynamical input that has to be provided by our calculations, but otherwise it is a natural reflection of the underlying symmetry, which may also explain the universality featured in (\ref{eq:main_result}). On the other hand, at the level of actual computations we have not found simple organizing principles making the universality transparent -- the agreement between $\ell=1$ and scalar contributions currently appears as a result of many cancellations among terms violating (\ref{eq:main_result}). For the case of multi-local channels, the structure of the computations becomes more obscure: (i) the contribution to the mutual information itself would be more theory-dependent, e.g. depending on the OPE coefficient $C_{OOO}$; (ii) the shape dependence involves higher-point correlation functions $\sim \langle T_{\mu\nu}\mathcal{O}...\mathcal{O}\rangle$ that cannot be fixed by symmetries; (iii) it is not clear how much of the shape-dependence can be constrained by consistency with (\ref{eq:special_def}), i.e. conformal ward identity. Putting these difficulties and uncertainties aside, if the result is truly universal, this may point to more hidden structures in the perturbative expansion of the entanglement first law governing the form of linear responses. Therefore, in the future it is interesting to further study the validity of, or otherwise modifications to (\ref{eq:main_result}) from more generic contributions, e.g. those listed in (\ref{item:check1}-\ref{item:check3}). 
    
    \item \textit{\textbf{Beyond the OPE limit: finite $R_B$ corrections}}
    
    In this paper, we focused on the OPE limit of mutual information. In the conformal frame of our interest, this corresponds to the point-like limit: $R_B\ll L,R_A$. The result (\ref{eq:main_result}) pertains only to the leading order contribution in this limit. In the future, it is important to extend the result beyond the OPE limit. This could be achieved by re-summing over the responses from all OPE contributions. Similar to ordinary OPE expansion, these contributions can be organized into conformal blocks associated with particular primary operators in the orbifold CFT (e.g. bi-local or multi-local channels in \ref{item:check1}-\ref{item:check3}). They include corrections from descendent operators, which are essentially derivative expansions that restore the full shape dependence on sub-region $B$. As a result, studying the shape dependence of mutual information at the level of conformal blocks will be an important future direction to pursue that can systematically push beyond the OPE limit \cite{future_work}. With the possibility of universal multiplicative form factor in front of each OPE contribution, it is plausible that after re-summation over descendent corrections, such feature survives and the linear response from each conformal block contribution is still encoded in a multiplicative form factor. In fact, the symmetry of the set-up is independent of the OPE limit -- the subregion-preserving symmetry group $\hat{\mathcal{R}}_{AB}$ exists for arbitrary $R_A, R_B$. Therefore, if the unversality and extremality are also valid beyond the OPE limit, the susceptibility for finite $R_B$ is simply given by: 
    \be
    \frac{\delta I_{AB}}{\sqrt{h(\Omega)}\delta \zeta(\Omega)} \propto G'(\alpha)\Big(1+G(\alpha)^2\Big)^{\frac{1-d}{2}} \left[\frac{G(\alpha)}{\sin{(\alpha)}}\right]^{d-3}
    \ee
  where $G(\alpha)$ is the complicated function of $\alpha$ that appears on the right-hand side of the first line in (\ref{eq:angular_map_full}). In the future, it is interesting to check if this is indeed the case. 
    
    \item \textit{\textbf{Higher order responses}}
    
    Last but not least, it is also important to extend our analysis of shape dependence beyond the linear order response in shape deformation. A natural next-step towards this is to compute the non-local quadratic order response \cite{future_work}: 
    \be
    \frac{\delta^2 I_{AB}}{\delta \zeta(x_1)\zeta(x_2)}, \;\; x_1 \neq x_2
    \ee
    An analogous quantity for the entanglement entropy on spheres (called entanglement density) were studied in \cite{Faulkner2016Apr}, and were found to encode universal data of the underlying CFTs. Similarly, we expect to learn more about CFT and its entanglement structures from a corresponding second order response of mutual information. For example, from a shape-dependence point of view, the strong sub-additivity constraint for mutual information \cite{Casini2021Sep} bares no content at linear order -- it is trivially saturated by the linear order responses due to its Markovian form. The constraint becomes meaningful only at the second order. In the face of extremality property of mutual information between spheres, we also need to push to the second order response to identify the nature of the extremum, i.e. is it a maximum or a minimum? It will be particularly interesting to study higher order responses in conjunction with higher order corrections in small $R_B$. We leave these for future investigations.  
    
\end{itemize}

\section*{Acknowledgments}
We thank Thomas Faulkner, Ling-Yan Hung, Shenghan Jiang, and Xinan Zhou for useful discussions. We thank Mark Mezei and Gonzalo Torroba for comments on the manuscript. This work is supported by National Science Foundation of China (NSFC) grant no. 12175238.

\appendix
\section{Computing the integrals}\label{sec:solve-the-integral}
 In this appendix, we will give some of the details  for computing the integrals  that we encounter when we try to calculate the shape dependence of mutual information. They take the following form:
 \begin{equation}
 	\begin{aligned}
 		\frac{\delta I_{A,B}}{\delta \xi(X^{i})} &= (i) \oint_{|z|=\tilde{\epsilon}} {d}z \,\  \left\langle  \bigg(T_{zz}(z,\bar{z}, X^{i}) + T_{z\bar{z}}(z,\bar{z}, X^{i})\bigg) \Delta \tilde{H}_{A, B} \right\rangle + h.c \\
 		&=	\mathcal{K}^{(1)} + \mathcal{K}^{(2)}
 	\end{aligned}
 \end{equation}
 where $\mathcal{K}^{(1)}, \mathcal{K}^{(2)}$ represent contributions from the single and double integral terms of $\Delta \tilde{H}_{A,B}$.  They take the explicit form:
 \begin{equation}
 	\begin{aligned}
 		\mathcal{K}^{(1)}&=   - i \rho^{2 \Delta} \oint_{|z|=\tilde{\epsilon}} {d} z \,\  \int_{-\infty}^{\infty} \frac{d s}{(2 \cosh(s/2))^{2\Delta +2}}   \bigg\langle \bigg(T_{zz}(z,\bar{z}, X^{i}) + T_{z\bar{z}}(z,\bar{z}, X^{i})\bigg)  \\
         & \times   \mathcal{O}\left(-X[s]\right) \mathcal{O}\left(X[0]\right) \bigg\rangle + h.c \\
 		\mathcal{K}^{(2)} &=  -\frac{\rho^{2 \Delta}
 		}{2 \pi}  \oint_{|z|=\tilde{\epsilon}} {d} z \,\ 
 		\int_{-\infty}^{\infty} \frac{d s_{j} d s_{k}}{4 \cosh^{2} (s_{j}/2)} \left( \frac{1}{e^{s_{k}+ i \epsilon}-1} + \frac{1}{e^{s_{k}+s_{j}}+1} \right) c_{1}(-is_{k} + \epsilon)  \\
 		&\times \left\langle \bigg(T_{zz}(z,\bar{z}, X^{i}) + T_{z\bar{z}}(z,\bar{z}, X^{i})\bigg) \mathcal{O}\left(-X[s_{j}+s_{k}]\right) \mathcal{O}\left(-X[s_{j}]\right) \right\rangle + h.c .
 	\end{aligned}
 \end{equation} 
where we recall that $X[s]^\pm= e^{\pm s}, X[s]^i=0$. For simplicity, we only focus on the $\oint {d z}$ integrals. The main task is to extract the residue of simple poles in $z$ from the integrands. Furthermore, using the analysis to be presented below it can be checked that only the correlators involving $T_{zz}$ can give rise to such poles of $z$, so we shall omit writing terms involving other components of stress tensor from now to declutter notations. The three-point functions $\left\langle T_{zz}OO\right\rangle$ by themselves do not exhibit simple poles in $z$. Similar to (\ref{single-interval-deform}) such poles can only emerge as a result of performing the modular integrals, i.e. we should identify the following behaviors from the integrands: 
\bea
\mathcal{F}^{(1)}(z,\bar{z}) &=& \int ds\; k_1(s) \left\langle T_{zz}\left(z,\bar{z},X^i\right)\mathcal{O}\left(-X[s]\right) \mathcal{O}\left(X[0]\right)\right\rangle \sim \frac{\sharp}{z} + ...\nonumber\\
\mathcal{F}^{(2)}(z,\bar{z}) &=&\int ds_j ds_k\; k_2\left(s_j,s_k\right) \left\langle T_{zz}\left(z,\bar{z},X^i\right)\mathcal{O}\left(-X[s_j+s_k]\right) \mathcal{O}\left(-X[s_j]\right)\right\rangle \sim \frac{\sharp}{z} +...\nonumber
\eea 
for the terms of $\mathcal{K}^{(1)}$ and $\mathcal{K}^{(2)}$ respectively. Doing these integrals exactly is neither pleasant nor illuminating. However, since we are only interested in the singularity of $\mathcal{F}^{(1,2)}(z,\bar{z})$ as $z,\bar{z}\to 0$, it is revealing to study their nature of divergences when setting $z,\bar{z}$ to be exactly zero. For example, a necessary condition for the emergence of simple pole is that $\mathcal{F}^{(1,2)}(0,0)$ be divergent. This will rule out the possibility of finding poles in $\mathcal{F}^{(1)}(z,\bar{z})$ which implies that $\mathcal{K}^{(1)}=0$. To check this, let us quote the three-point functions (\ref{TOO}) and compute $\mathcal{F}^{(1)}(0,0)<\infty$ explicitly as:
\bea
 \mathcal{F}^{(1)}(0,0) &=& \int^{\infty}_{-\infty} \frac{ds}{\left(2\cosh{(s/2)}\right)^{2\Delta+2}}\left\langle  T_{zz}(0,0; X^{i})  \mathcal{O}(-X[s]) \mathcal{O}(X[0])  \right\rangle \nonumber\\
 &=& -\frac{C_{12T}}{4 } \left(1+D\right)^{-(1+d)}\int^\infty_0 \frac{du\;u^{2\Delta-d/2}}{(u+1)^{4\Delta+3-d}},\;\;u\equiv e^{s} \nonumber\\
&=&-\left(\frac{C_{12T}\sqrt{\pi}}{2^{4\Delta-d+4}}\right)\left(1+D\right)^{-(1+d)}\frac{\Gamma\left(2\Delta-\frac{d}{2}+1\right)}{\Gamma\left(2\Delta-\frac{d}{2}+\frac{3}{2}\right)}<\infty
\eea
where we have denoted $D = \sum_{i=2}^{d-1} (X^{i})^{2} $.

So we are only left with the possible contribution from $\mathcal{K}^{(2)} $, whose integrand is given by quoting the corresponding three-point function as: 
\bea\label{eq:F2}
&&\mathcal{F}^{(2)}(z,\bar{z}) = \int^\infty_{-\infty} ds_j \int ^\infty_{-\infty}ds_k\; k_2\left(s_j,s_k\right)  \left\langle  T_{zz}(z,\bar{z}; X^{i})  \mathcal{O}\left(-X[s_j+s_k]\right) \mathcal{O}\left(-X[s_j]\right)  \right\rangle\nonumber \\
 &=& \frac{C_{12T}}{4 } \int^\infty_{-\infty} ds_k\int^{\infty}_{-\infty} ds_j \frac{k_2(s_j,s_k)\;  \left|w_{12}\right|^{d-2-2\Delta}\Big( w_{12} (\bar{z}-\bar{w}_{1}) (\bar{z}-\bar{w}_{2}) - \bar{w}_{12} D \Big)^{2}}{\Big((z-w_{1})(\bar{z} -\bar{w}_{1}) + D \Big)^{1+\frac{d}{2}} \Big((z-w_{2})(\bar{z} -\bar{w}_{2}) + D \Big)^{1+\frac{d}{2}}}\nonumber\nonumber\\
&=& (-1)^{\frac{d}{2}} \left(\frac{C_{12T}}{4 }\right)\int^\infty_0 dv\; (v-1)^{d-2}v^{1-\frac{d}{2}} \int^\infty_0 \frac{du}{(u+1)(uv+1)}\nonumber\\
&\times & \frac{\left(\bar{z}^2 u + (v^{-1}+1)\bar{z}+u^{-1}v^{-1}(1+D)\right)^2}{\left( uv \bar{z}+u^{-1}v^{-1}z + 1+D\right)^{1+\frac{d}{2}}\left( u \bar{z}+u^{-1}z + 1+D\right)^{1+\frac{d}{2}}}
\eea
where the coordinates $(w_{1,2},\bar{w}_{1,2})$ are given by $ w_{1} = - e^{s_{k}+s_{j}}, \bar{w}_{1} = -e^{-s_{k}-s_{j}}, w_{2} = - e^{s_{j}}, \bar{w}_{2} = - e^{-s_{j}}$ and $w_{12} = w_1-w_2$, etc. In the last step we have re-defined the modular parameters in terms of $ u = e^{s_{j}}, v=e^{s_{k}}$ and discarded the $|z|^2 = \tilde{\epsilon}^2\ll 1$ terms in the denominator. In this case, one can easily check that $\mathcal{F}^{(2)}(0,0)$ is divergent from the $u$ integral alone: 
\be\label{eq:singular}
\mathcal{F}^{(2)}(0,0) \propto \int^\infty_0 \frac{du\;u^{-2}v^{-2}}{(u+1)(uv+1)} (1+D)^{-d} \sim \lim_{u\to 0} \left(u^{-1}\right)=\infty
\ee
We can extract from the nature of divergence (\ref{eq:singular}) two pieces of information relevant for our purposes. Firstly the divergence is controlled by the $u\to 0$ limit of the integral; secondly the order of the divergence dictates that $\mathcal{F}^{(2)}(z,\bar{z})$ has a simple pole for $z$: 
\be 
\lim_{z\to 0} \mathcal{F}^{(2)}(z,\bar{z}) \to \frac{\sharp}{z}+...
\ee
This is because any finite value of $z$ provides an effective regulator for the $u\to 0$ divergence through the $u^{-1}v^{-1}z$ and $u^{-1}z$ terms in the denominator of (\ref{eq:F2}). Therefore we can focus on the $u\to 0$ regime of the $u$ integral and thus neglect the $\bar{z}$ terms in the denominator of (\ref{eq:F2}). Under this approximation, the relevant part of the integrand become a holomorphic function in $z$ \footnote{The treatment for the $\oint d\bar{z}$ integral of the correlators involving $T_{\bar{z}\bar{z}}$ proceeds in exactly analogous way in terms of anti-holomorphic functions.}. It is interesting to see that we do not need to complete both integral in $u$ and $v$ in order to reveal the simple poles in $z$, only the $ u $ integral alone is sufficient. The final result is then obtained by integrating the $v$-dependent residue over the remaining modular parameter $v$. By doing this we have partially switched the order of integration between $\oint dz$ and $\int dv$. As is discussed in section \ref{sec:discuss}, the fact that this works implies that the shape response exhibits additive behavior in the $v$ variable of the modular Hamiltonian, and may point to deeper aspects regarding the underlying structures of modular Hamiltonian in OPE. 

Now we lay out more details for the computations. We first focus on the $u$-integral of $\mathcal{F}^{(2)}$ and extract the residue of poles in $z$. We will be slightly more general than the specific form of (\ref{eq:F2}), since a similar procedure can be applied to computing contributions from spinning primary operator like $\mathcal{O}_\mu$ in section (\ref{sec:spin-1}), whose numerators from the three-point functions $\left\langle T_{zz} \mathcal{O}_\mu\mathcal{O}_\nu\right\rangle$ are more complicated functions of $u,v,z,\bar{z}$. Stripping off the $u$-independent factors, a generic $u$-integral term takes the following form: 
\bea\label{eq:u-int-1}
 &&\int_{0}^{\infty} {d} u \,\  \frac{(u+1)^{-1}( uv +1)^{-1} u^{d+n}}{\Big(u v + z/(1+D) \Big)^{1+d/2} \Big(u  + z/(1+D)\Big)^{1+d/2}},\;\;n\in \mathds{Z} \nonumber\\
 &=&   \frac{\Gamma(2+ d)}{\Gamma(1+d/2)^{2}} \int_{0}^{1} {d} w\;  w^{d/2}(1-w)^{d/2} \int_{0}^{\infty} \frac{du\;(u+1)^{-1}( uv +1)^{-1} u^{d+n}}{\Big(u v w + u(1-w)+ z/(1+D) \Big)^{2+d} }\;\;
\eea
where we have used the  Feymann parameter technique to further simplify the $u$ integral part. Now we can extract the singularity in $z\to 0$ of (\ref{eq:u-int-1}) by defining: 
\be 
 y= u +\delta, \;\;\delta = z(1+D)^{-1}(wv+1-w)^{-1}\ll 1
\ee
and zooming into the $u\sim \delta$ region of the $u$ integral in (\ref{eq:u-int-1}):
\bea\label{eq:u-int-2}
	&& \int^\infty_0  d{u} \,\ \frac{(u+1)^{-1}( uv +1)^{-1} u^{d+n}}{\Big(u (w v+1-w)+z/(1+D)\Big)^{d +2}}\nonumber\\
 	&=& \frac{1}{\left(w v +1-w\right)^{d+2}}\left(\int^\infty_\delta dy \,\ y^{-(d+2)}\left(y-\delta\right)^{d+n}\right)\left(1+ \mathcal{O}\left(\delta\right)\right)\nonumber \\
 	&=& \frac{1}{\left(w v+1-w\right)^{d+1+n}}\frac{\Gamma(1-n)\Gamma(d+1+n)}{\Gamma(d+2)}\left(\frac{z}{1+D}\right)^{-1+ n}\left(1+\mathcal{O}\left(\frac{z}{1 + D}\right)\right)
\eea
For the scalar contribution (\ref{eq:F2}) we only have one term with $n=0$, which corresponds to a simple pole in $z$. For the spin-1 contribution in (\ref{sec:spin-1}) there are terms with $u^n,n<0$, but they are multiplied additional powers of $z$ to still give at most simple poles in $z$ \footnote{If there are higher order poles in $z$, our result will receive contributions that depend on non-universal details of the cut-off tube, e.g. shape of the tube which are higher-order corrections in small $\epsilon$.}. Setting $n=0$ from now on, one can then collect the residue of the simple $z$-pole in (\ref{eq:u-int-2}), feed into the remaining $w$ integral in (\ref{eq:u-int-1}) and obtain  a hypergeometric function in $1-v$.  We shall not write down these intermediate details. Finally, one ends up with the following $v$-integral for $\mathcal{K}^{(2)}$:
 \begin{equation}
 	\begin{aligned}
 		\mathcal{K}^{(2)} &=   - \left(\frac{2 i \rho^{2 \Delta}}{d+1}\right)
 	 \left(\frac{C_{12T}}{4 } \right)  (1 + D)^{-d+1} (-1)^{d/2-1-2\Delta}  \\
 		& \times  \int_{0}^{\infty} {d} v  \,\   \frac{v^{2\Delta}}{(v-1)^{4\Delta+1-d}} \,\ {{}_{2}  F_{1}}\bigg(1+d, 1+\frac{d}{2} ; 2+d ; 1-v \bigg) \\
 	\end{aligned}
     \label{the-DI-term}
 \end{equation}
The hypergeometric function in the integrand admits a finite expansion for integer values of $d$. For example, in the case of $ d= 4$, it takes the form:
 \bea
{{}_{2}  F_{1}}\Big(1+d, 1+\frac{d}{2} ; 2+d ; 1-v \Big)  = \frac{20 v^{-1}}{(v-1)^{5}}-\frac{5 v^{-2}}{2(v-1)^{5}}-\frac{20 v}{(v-1)^{5}}+\frac{5 v^{2}}{2(v-1)^{5}}+\frac{30 \log {v}}{(v-1)^{5}}\nonumber
 \eea
The first four terms cancel in the integral due to the invariance of integrand under $v\to v^{-1}$, so we only need to keep the $\log{v}$ term. This pattern turns out to be valid for general even integer $ d $, whose $\log{v}$ terms are given by:
 \begin{equation}
 	\begin{aligned}
 		{{}_{2}  F_{1}}\Big(1+d, 1+\frac{d}{2} ; 2+d ; 1-v \Big)  &= (-1)^{d/2} \frac{\Gamma(2+ d)}{\Gamma(1+d/2)^{2}} \frac{\log (v)}{(v-1)^{d+1}} + \cdots
 	\end{aligned}
 \end{equation}
Plugging this into the $v$ integral,  we eventually obtain that:
 \begin{equation}
 	\begin{aligned}
 		\mathcal{K}^{(2)} &=   2  \Delta \rho^{2 \Delta}  \frac{\sqrt{\pi}}{4^{2\Delta + 1}} \frac{\Gamma(2\Delta+1)}{\Gamma(2\Delta + 3/2)}  \frac{ 2^{d-2} \Gamma\left(\frac{d-1}{2}\right) }{\pi^{d/2-1 / 2}}   (1 + D)^{-d+1} .  \\
 	\end{aligned}
 \end{equation}
For odd integer $d$ the analysis is slightly more involved, but in the end one obtains the same dependence on $d$, and we shall spare the details.

\bibliographystyle{JHEP} 
\bibliography{./references/ref}

\end{document}